\documentclass[twocolumn]{aastex631}
\usepackage{amsmath,amsfonts,amssymb,graphicx,chngcntr,multirow, float,booktabs, afterpage}
\usepackage[]{hyperref}
\hypersetup{colorlinks=true}

\newcommand{\rg}{R_g}
\newcommand{\rin}{R_{\rm in}}
\newcommand{\rout}{R_{\rm out}}

\newcommand{\Tprec}{T_{\rm prec}}
\newcommand{\Oprec}{\Omega_{\rm prec}}
\newcommand{\wLT}{\omega_{\rm LT}}
\newcommand{\Msun}{M_\odot}
\newcommand{\Mdot}{\dot{M}}
\newcommand{\Mdotfb}{\dot{M}_{\rm fb}}

\received{}
\revised{}
\accepted{}

\shorttitle{The chirp of TDE disks}

\begin{document}

\title{ The chirping of Lense-Thirring precession in tidal disruption event accretion flows }

\author{Andrew Mummery}
\affiliation{School of Natural Sciences, Institute for Advanced Study, 1 Einstein Drive, Princeton, NJ 08540, USA}
\author{Eliot Quataert}
\affiliation{Department of Astrophysical Sciences, Princeton University, Princeton, NJ 08544, USA}
\begin{abstract} 
Tidal disruption event X-ray light curves do not appear to show clear signs of periodic modulation. This is naively somewhat surprising 
since the stars that are disrupted originate at large scales in the galaxy where they cannot know about the orientation of the black hole spin axis.  This should lead to the formation of a misaligned disk that precesses due to Lense-Thirring torques, modulating X-ray emission from the inner disk regions. We argue that in fact the properties which are required for solid-body precession, namely a thick $(H/R\sim {\cal O}(1))$ disk, naturally lead to rapid precession period change, with a per-period increase of $\Delta T_{\rm prec}/T_{\rm prec} \sim {\cal O}(1-10)$, {as the disk spreads to larger radii to conserve angular momentum}. In other words the precession of thick TDE disks is aggressively chirped, washing out any possibility of observing multiple cycles other than for fine tuned regions of parameter space.  
The global disk alignment timescale is equally strongly chirped by the exact same mechanism, meaning that TDE disks will not in general align with the black hole spin axis during a super-Eddington phase, and should generically show global quasi-steady warp profiles at the beginning of any thin disk phase.   These results have important implications for interpreting timing features  associated with TDE disks, including models for quasi-periodic X-ray eruption timing phenomenology, and the observational properties of TDE disks in the initial transition to the thin disk phase. 
\end{abstract}
\keywords{
Accretion (14);
Supermassive black holes (1663);
X-ray transient sources (1852); 
Time domain astronomy (2109);
Tidal disruption (1696)
}

\section{Introduction}


The tidal disruption and subsequent accretion of unfortunate stars which are scattered onto near radial orbits about their galactic center black hole lead to bright multi-wavelength flares for which a couple of hundred sources are now known \citep[e.g.,][]{Yao2023,Guolo24, Grotova25, MummeryVV25}. Among a wide range of interesting astrophysical questions that these objects may offer insight into, they seem likely to be particularly pertinent probes of the physics and observational properties of misaligned accretion flows:  the stars  that are disrupted originate at large scales in the galaxy where they cannot know about the orientation of the black holes spin axis, and so their angular momentum vector should be  roughly $\sim$ isotropically distributed with respect to the black holes spin axis (this is only approximately true, as tidal forces in general relativity do depend on the relative angular momentum  orientation of the black hole and star, which will induce a bias at high black hole masses, see \citealt{Mummery24}). A disk whose normal is misaligned with the black holes spin axis is impacted by various relativistic effects \citep{LenseThirring1918, Bardeen75}, and it appears completely generic that a randomly sampled TDE disk will show a large angle of initial misalignment, and a probe of strong field gravity effects. 

Perhaps the most natural expectation for an observational signature of a misaligned accretion flow in a TDE is the observation of quasi-periodic modulation in its X-ray emission \citep[an argument which goes back to][with a detailed treatment following by \citealt{Franchini2016}]{StoneLoeb2012}. The reason being that this misaligned accretion flow is subject to Lense-Thirring torques which, as the disk is plausibly expected to be thick and possibly super-Eddington at early times, will be propagated in a wave-like manner across the entire disk \citep[e.g.,][]{Ogilvie1999, LubowOgilviePringle2002}. In such a limit the disk is expected to precess like a single solid body (i.e., with a  single characteristic frequency). The observers line of sight into the inner disk (which is known to be the origin of the thermal X-ray emission seen from TDE disks \citealt{Mummery_Wevers_23, Guolo24, Guolo26}) then rocks back and forth on this timescale, likely leading to clear observational signatures in emission sourced in these small scales (i.e., X-rays) and a relatively coherent periodicity. 

Broadly speaking, a clear signal of this effect has never been observed (we discuss existing, though ambiguous, observational claims later). This is despite the large (order 100) population of X-ray bright TDEs, with no source showing more than $\sim 2-3$ cycles which could be plausibly periodic. At first, this appears somewhat surprising. 

In this paper we argue that this is actually a very natural state of affairs, and that observing precession and periodic modulation in a TDEs X-ray light curve is in fact highly fine tuned, and not generic. The reason for this is that the precession frequency 
\begin{equation}
    \Omega_{\rm prec} \sim {\tau_\bullet \over J_{\rm disk}}, 
\end{equation}
where $\tau_\bullet$ is the torque provided by the black hole on the disk,  and $J_{\rm disk}$ is the angular momentum of the disk, is very sensitive to the distribution of angular momentum in the disk.  As the Lense-Thirring torque falls off rapidly with radius $\omega_{\rm LT} \sim R^{-3}$ (it is a relativistic effect driven by the spin), what is relevant is the fraction of the total angular momentum of the disk near its inner edge, i.e., 
\begin{equation}
    \Omega_{\rm prec} \sim {J_{\rm in} \omega_{\rm LT, {\rm in}} \over J_{\rm out}} \sim f_{J} \, \omega_{\rm LT, \, in}. 
\end{equation}
The purpose of this paper is to examine, effectively, the temporal evolution of this inner disk angular momentum fraction $f_{J}$. We will show that it is strongly time dependent, and indeed that it drops rapidly once the solid-body precession requirement of $H/R \sim 1$ is enforced on the flow. This leads to a rapidly shrinking precession frequency, or equivalently a rapidly growing (and therefore unobservable) precession period.  The base physical mechanism which produces this result is incredibly simple, for a disk to feel any torque at all some material must get down to the very innermost regions $R \sim R_I$ (the innermost stable circular orbit will be where the surface density peaks), but to do so having started at the circularisation radius it must have given its angular momentum to other fluid elements, pushing them further away. As the specific angular momentum of the fluid goes like $\ell \propto \sqrt{R}$ every $\delta m$ of disk material that gets from the circularisation radius down to $R\approx 0$ must have sent another $\delta m$ four times further away. If $H/R\sim 1$, then material spends very little time at small radii (it is accreting rapidly), but all the $\delta m$'s that are pushed outwards hang around, and a significant bank of angular momentum gets pushed to large radii, causing $f_J \to 0$, leading to $1/T_{\rm prec}\to 0$.

In effect, by construction, the precession of thick TDE disks (if they are formed) would be heavily chirped, and thus periodic modulation in X-ray emission will be unobservable for the vast majority of parameter space. This has important implications for the interpretation of timing signals of transients associated with tidal disruption event disks, including quasi periodic X-ray eruptions. 

An analogous process impacts the disk alignment timescale, similarly sending $1/t_{\rm align} \to 0$. This leads to the general prediction that every TDE disk exits a super-Eddington phase (if such a phase exists) with disk angular momentum misaligned with the black hole spin axis (i.e., TDE disks are generically warped as they enter the thin disk phase).  {In this paper we focus solely on this initial super-Eddington phase of accretion and leave an analysis of the subsequent thin disk phase to future work.   We also acknowledge upfront that the formation of the disk in the super-Eddington phase is itself uncertain because inefficient cooling during super-Eddington fallback can inhibit the formation of a compact disk (e.g., \citealt{Loeb1997,lu_bonnerot2020,Metzger22}).  The simple super-Eddington disk considered here ignores these complications and is thus  likely the `best case' set of assumptions for coherent disk precession; and yet we shall see that it is still very unlikely even in this model.}

The layout of this paper is the following, in the following section we lay out the base expressions, before examining the precession evolution in section \ref{sec:DeltaT}. We move to a numerical framework in section \ref{sec:num}. We examine the disk alignment process in section \ref{sec:align}, before discussing the implications of our results more broadly and concluding in section \ref{sec:con}. 

\section{Elementary Analysis}
In this and the following section we estimate the degree to which the precession period of a TDE disk ``wanders'' during a single precession cycle. This is intended to be an elementary scaling analysis, and as such we work entirely in the Newtonian limit. A numerical approach using full relativistic expressions will follow in section \ref{sec:num}. 

Consider a disk with a simple surface density profile $\Sigma(R) \propto R^{-p}$, extending from $\rin$ to $\rout$.  The Keplerian specific angular momentum is $\ell(R) = \sqrt{G M_\bullet R}$, so the angular momentum per unit radius is
\begin{equation}
    \frac{{\rm d}J}{{\rm d}R} = 2\pi R\Sigma(R)\ell(R) \propto R^{3/2-p} .
\end{equation}
A disk which has an angular momentum which is not aligned with the black holes spin axis undergoes nodal precession, driven by the  Lense-Thirring effect. In the Newtonian (i.e., leading order) limit this differential nodal precession occurs at a rate
\begin{equation}\label{eq:wLT}
    \wLT(R) = \frac{2aG^2 M_\bullet^2}{c^3 R^3} .
\end{equation}
If warp communication across the disk is faster than the differential precession  (known as the rigid-body regime) then the entire disk precesses at the angular-momentum-weighted average frequency
\begin{equation}\label{eq:Oprec_def}
    \Oprec = \frac{\displaystyle\int_{\rin}^{\rout} \wLT(R)\frac{{\rm d}J}{{\rm d}R}{\rm d}R}
  {\displaystyle\int_{\rin}^{\rout} \frac{{\rm d}J}{{\rm d}R}{\rm d}R} .
\end{equation}
This is an elementary integral in the Newtonian limit, we define $\epsilon \equiv \sqrt{\rin/\rout}$, so that $\epsilon \ll 1$ for an extended disk.  The  precession period $\Tprec = 2\pi/\Oprec$ follows trivially 
\begin{multline}\label{eq:Tprec}
\Tprec = \frac{\pi c^3}{aG^2 M_\bullet^2}
    \rout^{(5-2p)/2}\rin^{(1+2p)/2} \\
    \times     \left(\frac{1 + 2p}{5 - 2p}\right)
    \left(\frac{1 - \epsilon^{5-2p}}{1 - \epsilon^{1+2p}}\right)
\end{multline}
For $\epsilon \ll 1$ the correction factor approaches unity, and the net scaling is
\begin{equation}\label{eq:Tprec_scaling}
    \Tprec \propto \frac{1}{a}\rout^{(5-2p)/2}\rin^{(1+2p)/2} ,
\end{equation}
which can be readily understood. As $\wLT \propto R^{-3}$ is steeply falling, the torque is dominated by $\rin$ as essentially all the frame-dragging is exerted at the inner edge, while the moment of inertia of the disk lives at $\rout$.  The precession period is therefore set by two  separated scales
\begin{equation}\label{eq:Tprec_twoscale}
    \Tprec \sim t_{\rm LT}(\rin)\times\frac{{R_{\rm out}{\rm d}J/{\rm d R}|_{\rout}}}{R_{\rm in}{\rm d}J/{\rm d R}|_{\rin}} ,
\end{equation}
where $t_{\rm LT}(R) \equiv \pi c^3 R^3/(aG^2 M_\bullet^2) = 2\pi/\wLT(R)$ is the local (inner disk) Lense-Thirring period, and the second term is the inverse of the fraction of the total disk angular momentum (fixed in the case of a TDE to that of the incoming star) which resides near to $\rin$.

Note that the surface density normalisation cancels between the numerator and denominator of eq.~\eqref{eq:Oprec_def}, the precession frequency depends on the shape of the profile (through $p$, which is really just the distribution of the disks specific angular momentum) but not on its amplitude.  

A naive estimate of the periodic signal one might hope to observe in emission from a TDE disk would then follow from evaluating $T_{\rm prec}$ at the disk formation radius (an estimate for the initial disk outer edge), and comparing that to the bulk evolution timescale of the disk.

Following such logic, one uses the simple result that angular momentum conservation in the stellar debris implies that the disk circularizes at twice the tidal radius
\begin{equation}\label{eq:Rcirc}
    R_{\rm circ} \approx {2 R_T \over \beta} = {2\over \beta} R_\star \left(\frac{M_\bullet}{M_\star}\right)^{1/3} ,
\end{equation}
where we define the usual penetration factor of the stellar orbit $\beta$ (the depth within the tidal radius reached by the pericentre of the stellar orbit). In units of the gravitational radius $\rg \equiv GM_\bullet/c^2$,
\begin{equation}\label{eq:rcirc}
    \frac{R_{\rm circ}}{\rg}
    \approx {94 \over \beta} \left(\frac{M_\bullet}{10^6\Msun}\right)^{-2/3} \left(\frac{R_\star}{R_\odot}\right) \left(\frac{M_\star}{\Msun}\right)^{-1/3} .
\end{equation}
The orbital period at this fiducial radius $R_{\rm circ}$ is
\begin{equation}\label{eq:torb}
    t_{\rm orb}(R_{\rm circ}) = 2\pi \sqrt{\frac{R_{\rm circ}^3}{G M_\bullet}} = 2\pi \sqrt{\frac{8 R_\star^3}{\beta^3 G M_\star}} ,
\end{equation}
which evaluates to
\begin{equation}\label{eq:torb_num}
    t_{\rm orb}(R_{\rm circ}) \approx {0.33 \over \beta^{3/2}} \left(\frac{R_\star}{R_\odot}\right)^{3/2} \left(\frac{M_\star}{\Msun}\right)^{-1/2}\text{days} .
\end{equation}

At the moment of disk formation, $\rout \approx R_{\rm circ}$.  Evaluating eq.~\eqref{eq:Tprec} at this radius gives the initial precession period. For $p = 1/2$ (the natural profile for a thick disk, as we shall discuss later) and $\rout = R_{\rm circ}$, we neglect the disk size ratio $\epsilon_0 \equiv \sqrt{\rin/R_{\rm circ}} \ll 1$, and the precession period (eq.~\ref{eq:Tprec}) becomes
\begin{equation}\label{eq:Tprec_Rcirc_asymp}
    \Tprec(R_{\rm circ}) \approx \frac{\pi c^3}{2aG^2 M_\bullet^2} R_{\rm circ}^{2}\rin.
\end{equation}
Here $R_{\rm circ}$ depends on $(M_\bullet, M_\star, R_\star)$ through eq.~\eqref{eq:Rcirc}, and $\rin = r_{I}(a)GM_\bullet/c^2$ where $r_{I}$ is the standard dimensionless ISCO radius.  The result is a function of $(M_\bullet, a, M_\star, R_\star, \beta)$ alone, and evaluates to 
\begin{equation}\label{eq:Tprec_normalised}
    \Tprec(R_{\rm circ}) \approx \left(\frac{{r_{I} r_\star^2  }}{a M_6^{1/3} m_\star^{2/3} \beta^2} \right) \times 1\, \text{day} ,
\end{equation}
where we have introduced the dimensionless stellar and black hole parameters
\begin{equation}\label{eq:normalisations}
    M_6 \equiv \frac{M_\bullet}{10^6\Msun} ,\qquad
    m_\star \equiv \frac{M_\star}{\Msun} ,\qquad
    r_\star \equiv \frac{R_\star}{R_\odot} .
\end{equation}
A naive estimate of the evolutionary time of the rise to peak of a TDE's X-ray flare would be the fallback time (the orbital period of the most-bound debris), which is
\begin{multline}\label{eq:tfb}
    t_{\rm fb} = \frac{\pi}{\sqrt{2}} \left(\frac{M_\bullet}{M_\star}\right)^{1/2} \sqrt{\frac{R_\star^3}{G M_\star}}
\\     \approx 41 M_6^{1/2} r_\star^{3/2} m_\star^{-1}\,\,\text{days} .
\end{multline}
One would conclude that a $\sim 10\%$ duty cycle periodicity in X-ray emission would be simple enough to observe \citep[e.g.,][]{StoneLoeb2012, Franchini2016}. An obvious question then becomes why this is not always seen in TDE X-ray light curves. 

\section{Period evolution}\label{sec:DeltaT}
We will assume that our TDE disk is fed material at a rate $\dot M_{\rm fb}(t)$ at the circularization radius, and that matter is also accreted onto the black hole at a rate $\dot M_{\rm acc}(t)$, or in other words the total disk mass evolves via 
\begin{equation}\label{eq:Md_dot_general}
\dot M_d = \dot M_{\rm fb} - \dot M_{\rm acc} .
\end{equation}
Accompanying this matter is angular momentum, and the rate of change of disk angular momentum is
\begin{equation}\label{eq:Jd_dot_general}
    \dot{J}_d = \Mdotfb\ell_{\rm circ} - \Mdot_{\rm acc}\ell_{\rm in} .
\end{equation}
We shall define the specific angular momentum ratio
\begin{equation}\label{eq:q_def}
    q = \frac{\ell_{\rm in}}{\ell_{\rm circ}} = \sqrt{\frac{R_{\rm in}}{R_{\rm circ}}} ,
\end{equation}
where $\ell_{\rm circ} = \sqrt{G M_\bullet R_{\rm circ}}$ and $\ell_{\rm in} = \sqrt{G M_\bullet R_{\rm in}}$.
Since $\ell_{\rm in} = q\ell_{\rm circ} \ll \ell_{\rm circ}$, each unit mass passing through the disk deposits an angular momentum $\sim \ell_{\rm circ}(1 - q)$. Angular momentum therefore accumulates with time. 

Since $\Tprec \propto R_{\rm out}^{(5-2p)/2}$ (eq.~\ref{eq:Tprec_scaling}) and $R_{\rm out} \propto \langle \ell \rangle^2$ (from the fact that the specific angular momentum of the flow is maximized in the outer regions $\langle \ell \rangle \propto \sqrt{GM_\bullet R_{\rm out}}$), the precession period tracks the average specific angular momentum of the disk
\begin{equation}\label{eq:Tprec_ell}
    \Tprec \propto \langle \ell \rangle^{5-2p} ,\qquad \langle \ell \rangle \equiv \frac{J_d}{M_d} .
\end{equation}
Taking the logarithmic derivative of eq.~\eqref{eq:Tprec_ell} leads to 
\begin{equation}\label{eq:dlnTprec}
    \frac{\dot T_{\rm prec}}{\Tprec} = (5-2p)\frac{\dot{\langle \ell \rangle}}{\langle \ell \rangle} .
\end{equation}
Expanding the derivative of  $\langle \ell \rangle = J_d/M_d$ gives
\begin{equation}
    \dot{\langle \ell \rangle} = \frac{1}{M_d}\left[\dot{J}_d - \langle \ell \rangle\dot{M}_d\right] ,
\end{equation}
where $\dot{J}_d = \Mdotfb\ell_{\rm circ} - \Mdot_{\rm acc}\ell_{\rm in}$ (eq.~\ref{eq:Jd_dot_general}) and $\dot{M}_d = \Mdotfb - \Mdot_{\rm acc}$ (eq. \ref{eq:Md_dot_general}), or finally 
\begin{equation}\label{eq:Tdot_exact}
 \frac{\dot T_{\rm prec}}{\Tprec} = \frac{5-2p}{M_d\langle \ell \rangle}\left[\Mdotfb(\ell_{\rm circ} - \langle \ell \rangle) +\Mdot_{\rm acc}(\langle \ell \rangle - \ell_{\rm in})\right] .
\end{equation}
For a super-Eddington rigid-body precessing disk model to be realised, the viscous time of the flow 
\begin{equation}
t_{\rm visc} \approx \alpha^{-1} \left({H\over R}\right)^{-2} \Omega^{-1}_{\rm orb}(R_{\rm circ})  ,
\end{equation}
must be short by definition (as one needs $H/R\sim 1$). In this limit, the fallback time will be guaranteed to be long compared to the viscous time 
\begin{equation}
{t_{\rm fb}\over t_{\rm visc}} \approx \alpha \beta^{3/2}\left({H\over R}\right)^2 \left({M_\bullet\over M_\star}\right)^{1/2} \approx \alpha\beta^{3/2} \left({M_\bullet\over M_\star}\right)^{1/2}\gg 1.
\end{equation}
In the $t_{\rm fb}\gg t_{\rm visc}$ limit, one will find that the accretion rate closely tracks the fallback rate at early times $\dot M_{\rm acc} \approx \dot M_{\rm fb}$, and the initial disk mass will be low $M_d \approx \dot M_{\rm fb} t_{\rm visc}$. Initially, the specific angular momentum of the flow will be roughly that of the circularized angular momentum $\left\langle \ell \right \rangle \approx \ell_{\rm circ}$, and we have 
\begin{equation}
 \frac{\dot T_{\rm prec}}{\Tprec} \approx \frac{(5-2p)(1-q) }{ t_{\rm visc}} ,
\end{equation}
or integrating over one period 
\begin{equation}\label{eq:DeltaT_earlytime}
\frac{\Delta \Tprec}{\Tprec} \approx (5-2p)(1-q)\frac{\Tprec(R_{\rm circ})}{t_{\rm visc}}
\end{equation}
For $\epsilon \ll 1$ the rigid-body precession period (eq.~\ref{eq:Tprec_Rcirc_asymp}) is
\begin{equation}\label{eq:Tprec_tLT}
    \Tprec(R_{\rm circ}) = \left(\frac{1+2p}{5-2p}\right) t_{\rm LT}(\rin)\left(\frac{R_{\rm circ}}{\rin}\right)^{(5-2p)/2} ,
\end{equation}
where $t_{\rm LT}(R) \equiv \pi c^3 R^3/(aG^2 M_\bullet^2)$ is the local Lense-Thirring precession period.  Dividing by the thick disk viscous time $t_{\rm visc} = t_{\rm orb}(R_{\rm circ})/\alpha (H/R)^2$ leaves 
\begin{equation}\label{eq:drift_explicit_p}
\frac{\Delta \Tprec}{\Tprec} \approx (1+2p)(1-q)\alpha  \left({H\over R}\right)^2 \frac{t_{\rm LT}(\rin)}{t_{\rm orb}(\rin)}\left(\frac{R_{\rm circ}}{\rin}\right)^{1-p},
\end{equation}
where the central factor 
\begin{equation}
{t_{\rm LT}(\rin)\over t_{\rm orb}(\rin)} = {r_{I}^{3/2}\over 2a},
\end{equation}
is the number of inner-edge orbits per local Lense-Thirring precession cycle --- a pure GR quantity, of order unity at high spin ($\approx 0.9$ for $a = 0.99$) and $\sim 10$ at $a = 0.5$. 

It looks at first like one could in principle have a relatively fixed period therefore by taking $p>1$. There are two reasons why this does not work as an argument. The first comes from assuming that it is indeed the case, and then examining the consequences.  If one takes  $p>1$ the one finds (equation \ref{eq:Tprec}) that the precession time grows more slowly than the outer orbital time as the disk expands. If this happens then quite rapidly   the disk enters a regime where it cannot precess, as the time to communicate a sound wave across the disk 
\begin{equation}
    t_{\rm wave} \sim R_{\rm out}/c_s \sim t_{\rm orb}(R_{\rm out}) (H/R)^{-1} \propto R_{\rm out}^{3/2} ,
\end{equation} 
will grow above the apparent precession time. Indeed, we have $T_{\rm prec}/t_{\rm wave} \propto R_{\rm out}^{1-p}$, meaning that $p> 1$ (which prevents rapid chirping) leads to a killing off of precession through an entirely different mechanism –– the flow cannot behave like a solid body on the required time frame, as the signal which must be communicated across the disk by pressure waves (that there is an inner disk torque which the flow should respond to) cannot propagate as quickly as the disk wants to precess.  In this limit the disk would stop precessing, and settle into a warped structure.

Secondly, this $p>1$ limit is again constrained to not be the natural case owing to the thick disk assumption, whereby the accretion velocity 
\begin{equation}
    v_R \approx \alpha \left({H\over R}\right)^2 v_K \approx \alpha \sqrt{GM_\bullet\over R}
\end{equation}
is forced to follow the Keplerian velocity when $H/R\sim 1$ is fixed, leading to 
\begin{equation}
    \Sigma = {\dot M \over 2\pi R v_R} \sim R^{-1/2}, 
\end{equation}
i.e., $p = 1/2$. In fact, all (thermally stable) $\alpha$-type disk models have $p<1$, a result which is only strengthened by the presence of super-Eddington winds (as we shall show shortly).

Taking $p=1/2$ we finally have
\begin{equation}\label{eq:drift_explicit}
\frac{\Delta \Tprec}{\Tprec} \approx \alpha \left({H\over R}\right)^2 {r_I \over a}\sqrt{\frac{R_{\rm circ}}{R_g}}, 
\end{equation}
or numerically 
\begin{equation}\label{eq:drift_explicit_2}
\frac{\Delta \Tprec}{\Tprec} \approx 10 \alpha \left({H\over R}\right)^2 {r_I\over a} M_6^{-1/3} r_\star^{1/2} m_\star^{-1/6} \beta^{-1/2} . 
\end{equation}
By construction, in a solid body precessing disk, one has $H/R \sim 1$, meaning that for canonical TDE parameters this period drift is $\sim {\cal O}(1-10)$, and one would not expect in reality to see a periodic signal in the data. 
\subsection{Super-Eddington winds}\label{sec:wind}

The above expressions assumed that mass leaves the disk only by accreting through the inner disk edge.  At super-Eddington accretion rates (required for a solid body precession), radiation-pressure-driven winds will likely eject a significant fraction of the inflowing mass from intermediate radii, before it has accreted.  This could in principle modify our conclusions, if a significant fraction of angular momentum was blown away, rather than accumulating in the disk. We will show that in fact a super-Eddington wind makes things worse (i.e., increases the period drift) owing to the impact of the wind on the disk surface density profile.  We will assume that such a radiation pressure driven wind imparts negligible azimuthal torque, likely a reasonable approximation,  so that each wind parcel carries only the specific angular momentum of the fluid element from which it was launched.

Let the wind remove mass at rate $\Mdot_w$ with average specific angular momentum $\ell_w$.  The budgets become
\begin{equation}\label{eq:Md_dot_wind}
    \dot{M}_d = \Mdotfb - \Mdot_{\rm in} - \Mdot_w ,
\end{equation}
\begin{equation}\label{eq:Jd_dot_wind}
    \dot{J}_d = \Mdotfb\ell_{\rm circ} - \Mdot_{\rm in}\ell_{\rm in} - \Mdot_w\ell_w .
\end{equation}
Running through the same $\dot{\langle \ell \rangle} = (\dot{J}_d - \langle \ell \rangle \dot{M}_d)/M_d$ algebra yields
\begin{multline}\label{eq:Tdot_wind}
    \frac{\dot T_{\rm prec}}{\Tprec} = \frac{5-2p}{M_d\langle \ell \rangle}\Big[\Mdotfb(\ell_{\rm circ} - \langle \ell \rangle) + \Mdot_{\rm in}(\langle \ell \rangle - \ell_{\rm in}) \\ + \Mdot_w(\langle \ell \rangle - \ell_w)\Big] .
\end{multline}
Without the wind, the mass that goes into $\Mdot_w$ would instead have eventually accreted through the ISCO carrying $\ell_{\rm in}$.  Defining the no-wind accretion rate $\Mdot_{\rm in,0} = \Mdot_{\rm in} + \Mdot_w$, the difference between the wind and no-wind drift rates is (this is simply a trivial manipulation)
\begin{equation}
    \left(\frac{\dot{T}}{T}\right)_{\rm wind} - \left(\frac{\dot{T}}{T}\right)_0 = \frac{5-2p}{M_d\langle \ell \rangle}\Mdot_w(\ell_{\rm in} - \ell_w) .
\end{equation}
Since $\ell_w > \ell_{\rm in}$ (the wind is launched from $R > \rin$), this is negative, and by removing angular momentum the wind does indeed slow the period evolution.  

However, the period evolution is impacted both by the angular momentum budget, and by the disk surface density profile (via $p$).  By removing matter as it propagates through the disk, a wind modifies the effective $p$ of the surface density profile, which works to counteract this angular momentum draining effect.

To estimate both the fraction of mass lost into winds $f_w\equiv \dot M_w/\dot M_{\rm fb}$ and the angular momentum taken away $\lambda_w \equiv \ell_w/\ell_{\rm circ}$ suppose that at each radius a fraction $\varepsilon$ of the local accretion rate is lost to a wind 
\begin{equation}\label{eq:wind_model}
    \frac{{\rm d}\Mdot_w}{{\rm d}\ln R} = \varepsilon\Mdot(R) ,
\end{equation}
which is effectively an assumption that the local wind efficiency is scale-free.  As the mass lost to winds must modify the local accretion rate, we have ${\rm d}\Mdot/{\rm d}\ln R = \varepsilon\Mdot$, giving the quasi-steady radial profile
\begin{equation}\label{eq:Mdot_wind_profile}
    \Mdot(R) = \Mdotfb\left(\frac{R}{R_{\rm circ}}\right)^\varepsilon ,
\end{equation}
since $\Mdot(R_{\rm circ}) = \Mdotfb$ by definition (the circularisation radius is where the fallback debris joins the disk).  We have explicitly assumed here that the disk has reached a quasi-steady state at each radius, which is in effect once again assuming that $t_{\rm visc} \ll t_{\rm fb}$ (which is consistent with the proposed super-Eddington condition that drives the wind and also guarantee $H/R \sim 1$ and hence short $t_{\rm visc}$).  The mass reaching the ISCO is
\begin{equation}\label{eq:Mdot_in_wind}
    \Mdot_{\rm acc} = \Mdotfb q^{2\varepsilon} ,
\end{equation}
and the wind fraction is $f_w = 1 - q^{2\varepsilon}$.

As the thick disk limit again forces a velocity scaling $v_R \simeq \alpha (H/R)^2 v_K \sim R^{-1/2}$, this modified local accretion rate leads to 
\begin{equation}
    \Sigma \propto R^{\varepsilon- 1/2},
\end{equation}
shifting $p$ to $1/2-\varepsilon$. 

In the early-time short-viscous-time limit ($\langle \ell \rangle \approx \ell_{\rm circ}$), the ratio of wind-modified to no-wind period drift is therefore modified both by the angular momentum removed from the system, but also by the change in density profile, resulting in a final 
\begin{equation}\label{eq:wind_suppression}
\frac{(\dot T/T)_{\rm wind}}
     {(\dot T/T)_{0}}
\;=\;
(1-\varepsilon)\,
\Bigl(\frac{\rout}{\rin}\Bigr)^{\!\varepsilon}\,
\frac{1-q^{\,4+2\varepsilon}}{1-q^{\,2-2\varepsilon}}\,
\bigl[\,1 - \lambda_{w}\,f_{w}\,\bigr],
\end{equation}
The mass-weighted specific angular momentum of the wind is
\begin{equation}
    \ell_w = \frac{1}{\Mdot_w}\int_{\rin}^{R_{\rm circ}} \ell(R)\frac{{\rm d}\Mdot_w}{{\rm d}R}{\rm d}R ,
\end{equation}
where $\ell(R) = \sqrt{GM_\bullet R}$ is the local Keplerian specific angular momentum and ${\rm d}\Mdot_w/{\rm d}R = \varepsilon\Mdot(R)/R$.  We can therefore compute $\ell_w$ via 
\begin{multline}
    \ell_w = \frac{\varepsilon\Mdotfb}{\Mdot_wR_{\rm circ}^\varepsilon}\int_{\rin}^{R_{\rm circ}} \sqrt{GM_\bullet}R^{\varepsilon - 1/2}{\rm d}R
        \\ = \left(\frac{\varepsilon}{\varepsilon + \tfrac{1}{2}}\right)\left(\frac{1 - q^{2\varepsilon + 1}}{1 - q^{2\varepsilon}}\right)\ell_{\rm circ} ,
\end{multline}
or for $q \ll 1$ 
\begin{equation}\label{eq:lambda_w}
    \lambda_w = \frac{2\varepsilon}{2\varepsilon + 1} .
\end{equation}

In the highly super-Eddington limit ($f_w \to 1$) with $q \ll 1$ (this second limit cannot be taken for $\varepsilon \to 1$), the suppression factor (eq.~\ref{eq:wind_suppression}) becomes
\begin{equation}\label{eq:wind_suppression_limit}
    \frac{(\dot{T}/T)_{\rm wind}}{(\dot{T}/T)_0} \approx \frac{1-\varepsilon}{2\varepsilon + 1} \left({\rout \over \rin}\right)^\varepsilon,
\end{equation}
The factor $\rout/\rin$ raised to the power $\varepsilon$ generally (for typical TDE parameters) beats the suppression pre-factor, leading to more aggressive chirping in the presence of winds.


\section{Numerical analysis}\label{sec:num}
The previous section used a number of simplifying assumptions to understand the broad parameter scaling of a typical TDE thick disk precession chirp. In this section we move to a numerical analysis of the evolving accretion flow to fully capture the evolution of the disk precession period. 

We shall assume that mass is fed into the accretion flow following the fallback rate at the circularisation radius, i.e., the disk evolution equations have a source term
\begin{equation}
    {\cal S}_{M}(r, t) = {M_\star \over 3t_{\rm fb}} \left({t \over t_{\rm fb}}\right)^{-5/3} \, \delta(r - R_{\rm circ}) \, \Theta(t-t_{\rm fb}),
\end{equation}
where the $\delta$-function is an approximation in the extreme limit of prompt circularisation, and the Heaviside $\Theta$ function enforces the fact that material only returns after one fallback time. 

The accretion disk surface density is then solved via the method of Greens functions, namely 
\begin{equation}
    \Sigma(r, t) = \int_0^\infty\int_0^t {\cal S}_M(r', t') {G}_\Sigma(r, t| r', t')\, {\rm d}t'\, {\rm d}r' ,
\end{equation}
where $G_\Sigma$ is the Greens function of the surface density. For our feeding prescription this becomes 
\begin{equation}
    \Sigma(r, t) = {M_\star \over 3t_{\rm fb}} \int_{t_{\rm fb}}^t \left({t' \over t_{\rm fb}}\right)^{-5/3} \, G_\Sigma(r, t| t', R_{\rm circ})\, {\rm d}t'. 
\end{equation}
We also compute the accretion rate across the ISCO, in an analogous fashion 
\begin{equation}
    \dot M(r_I, t) = {M_\star \over 3t_{\rm fb}} \int_{t_{\rm fb}}^t \left({t' \over t_{\rm fb}}\right)^{-5/3} \, G_{\dot M}(r_I, t| t', R_{\rm circ})\, {\rm d}t', 
\end{equation}
where $G_{\dot M}$ is the mass accretion rate Greens function. For both $G_\Sigma$ and $G_{\dot M}$ we use the \cite{Mummery23a} relativistic Green's function solutions, which include the effects of black hole spin in the disk fluid dynamics. We take the particular Greens function corresponding to a fixed aspect ratio of the disk (i.e., one which leads to $p=1/2$). This Greens function is also the solution which best reproduces the time-evolution of full 3D GRMHD simulations in the thick disk limit (Mummery et al., 2026, submitted).  We set the viscous time with a simple $\alpha$ parameter under the assumption that $H/R=1$ (i.e., we force a thick disk limit by construction) 
\begin{equation}
    t_{\rm visc} = {1 \over \alpha} \sqrt{R_{\rm circ}^3 \over G M_\bullet}. 
\end{equation}
We again stress that this viscous time parameterisation is a choice we make simply to try and force the best possible case for disk precession (i.e., by assuming the disk is thick), we do not claim that this is the actual limit that will be reached by a realistic TDE disk. 

Note that our analysis of period chirping is only dependent on the product $\alpha (H/R)^2$, and so the value of $\alpha$ in our simulations is really an effective parameter $\alpha \equiv \alpha_{\rm true} (H/R)_{\rm true}^2$. We fix $\alpha = 0.01$ while making plots as this value of the effective product is at the faster evolution end seen in TDE X-ray light curves \citep[namely the source AT2019dsg, which shows the most rapid X-ray evolution of known sources][]{mummery2024fitted, Guolo25time}. We anticipate a super-Eddington flow to be at the fastest end of the known distribution of TDE viscous times. We note that the GRMHD simulations of \cite{Guo25} showed a slightly higher effective parameter of $\alpha \sim 0.05$, but that this was an average over long timescales. 

From the surface density solution we compute the disks precession period from 
\begin{equation}
    T_{\rm prec}(t) = {2\pi \over \left\langle \Omega_{\rm prec}\right\rangle} = {2\pi \int r U_\phi(r, a) \Sigma(r, t)\, {\rm d}r \over \int r U_\phi(r, a) \Sigma(r, t) \Omega_{\rm LT} (r, a) \, {\rm d}r},
\end{equation}
where we use the full relativistic test particle solutions 
\begin{align}
   \Omega_\phi(r,a)   &= \frac{c^{3}}{GM_\bullet}\,\frac{1}{r^{3/2}+a},\\[2pt]
   \Omega_z(r,a)   &= \Omega_\phi(r,a)\,
                 \sqrt{1 - \tfrac{4a}{r^{3/2}} + \tfrac{3a^{2}}{r^{2}}},\\[2pt]
   \Omega_{\rm LT}(r,a)   &= \Omega_{\phi}(r,a) - \Omega_z(r,a),\\[2pt]
   U_\phi(r,a) &= \frac{G M_\bullet}{c}\,
                 \frac{r^{2} - 2a\sqrt{r} + a^{2}}
                      {r^{3/2}\sqrt{1 - 3/r + 2a\,r^{-3/2}}},
\end{align}
where in these expressions $r$ and $a$ are dimensionless.

As a way to visualize the chirp of the TDE disk concretely, we define the Lense-Thirring phase by 
\begin{equation}
    \phi(t) \equiv 2\pi\int_0^t {{\rm d} t' \over T_{\rm prec}(t')},
\end{equation}
and then plot a Lense-Thirring amplitude which is given simply by $A(t) = \sin\phi(t)$. One might expect observable luminosities to vary with $A(t)$, as the view to the inner disk sweeps in and out of the observers line of sight.

\begin{figure*}
    \centering
    \includegraphics[width=0.95\linewidth]{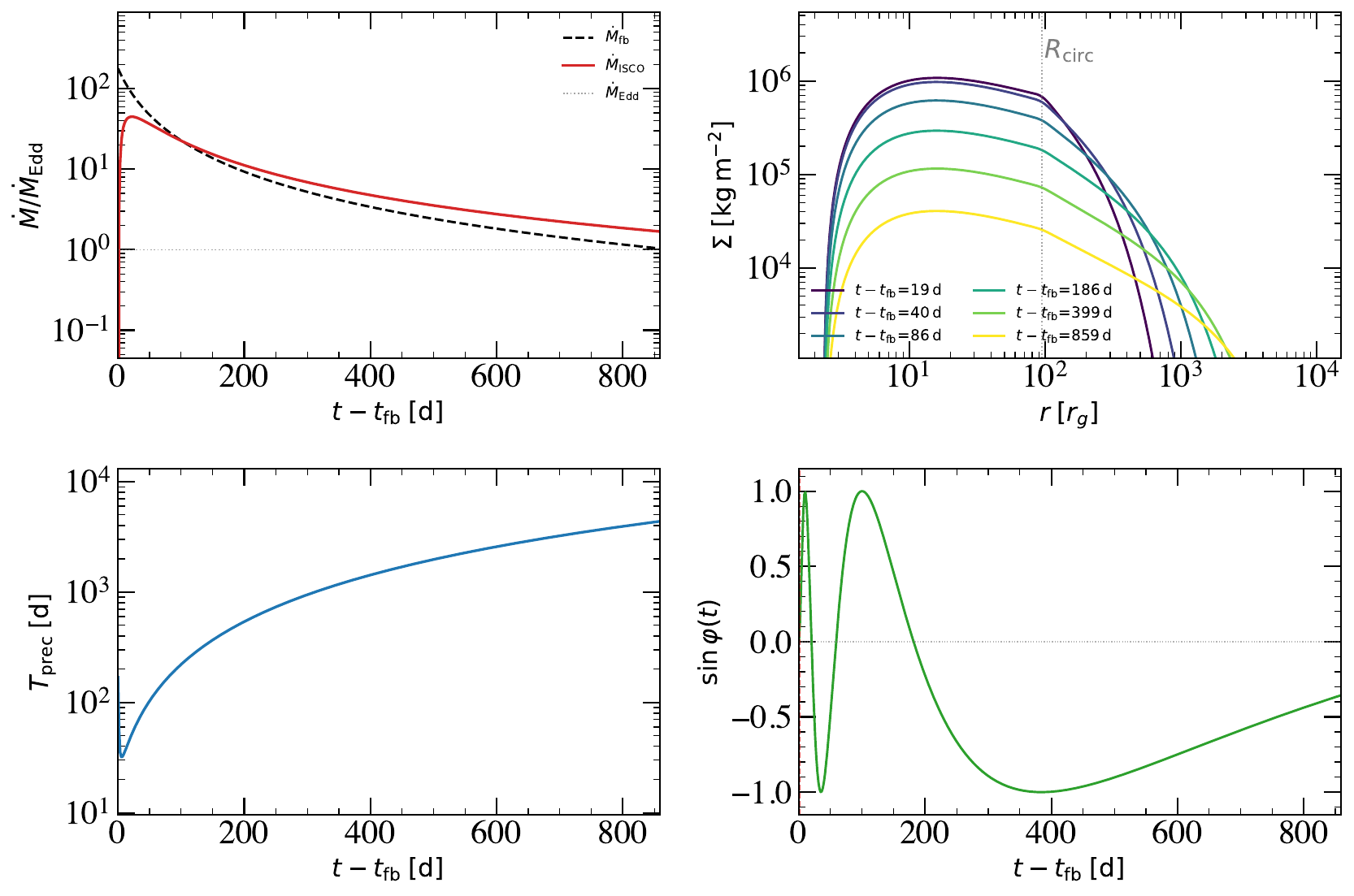}
    \caption{An example of the evolving precession period and chirp profile of a thick TDE accretion flow. We use a canonical set of disk and black hole parameters, namely $\alpha = 0.01, M_\bullet = 10^6 M_\odot, M_\star = M_\odot, a =0.9$ and $\beta = 1$. We set the viscous timescale by forcing $H/R = 1$ in the solution. By construction (as the viscous time is short), the accretion rate onto the black hole (red solid curve) closely tracks the fallback rate (black dashed curve), shown in the upper left panel. However, this prompt accretion forces rapid disk spreading (upper right) and associated rapid precession period evolution (lower left). This heavily chirps the Lense-Thirring phase (lower right), preventing any chance of seeing clear periodic signals.  }
    \label{fig:canonical}
\end{figure*}

As an example of the properties of these solutions, we show in Figure \ref{fig:canonical} a canonical set of disk parameters $\alpha = 0.01, M_\bullet = 10^6 M_\odot, M_\star = M_\odot, a =0.9$ and $\beta = 1$.  Note that the canonical precession period estimate in this limit is $T_{\rm prec}\sim 1$ day.  In Figure \ref{fig:canonical} we see that, by construction, the accretion rate onto the black hole tracks the fallback rate closely (top left panel), but that other than that, none of the other properties of the flow follow the conventional precessing disk picture. The surface density of the disk rapidly spreads (top right panel), which means that the precession timescale of the disk never even gets close to the value one would estimate from a disk extending from the ISCO out to the circularisation radius. This is because before any material has made it down to the ISCO the flow has already spread by a factor of a few (simply because the flow must expand even before any flow has made it down to the horizon, as angular momentum is redistributed). 

As the flow then evolves, the precession period shoots up (lower left panel), leading to a Lense-Thirring amplitude that undergoes $\sim 1$ full cycle in the super-Eddington phase (as opposed to the $\sim 800$ expected naively). 

\begin{figure*}
    \centering
    \includegraphics[width=0.95\linewidth]{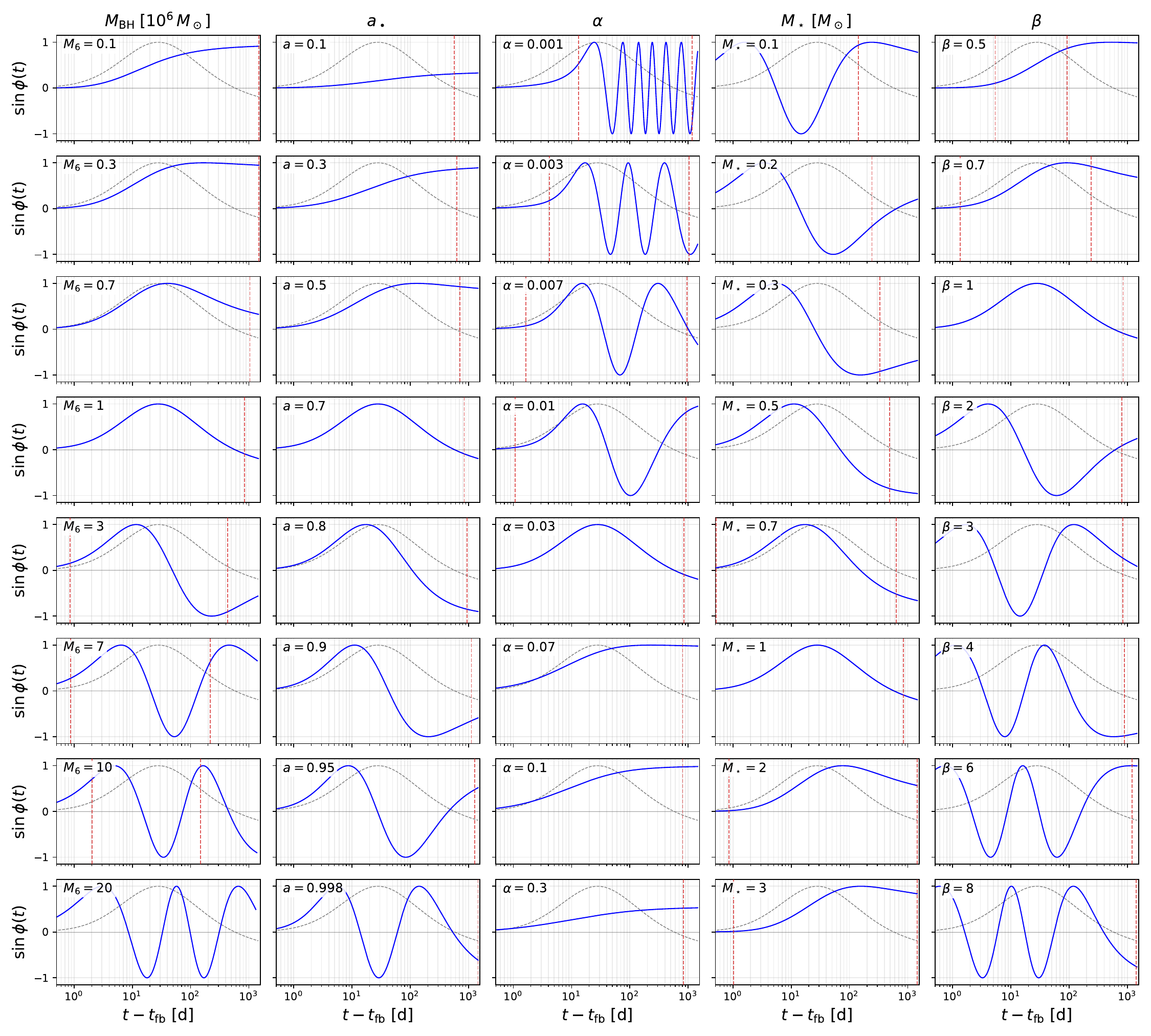}
    \caption{A library of the chirp profiles of thick ($H/R = 1$ forced throughout) TDE disks, for a range of different values of the parameters $M_\bullet, a, M_\star, \alpha$ and $\beta$. The ``chirp profile'' is the  sine of the Lense-Thirring phase $\phi(t)$. The time axis is displayed from the time of first material fallback on a logarithmic scale. The red dashed vertical lines denote the time frame over which the ISCO accretion rate is super-Eddington. For the vast majority of phase space one is unlikely to see a complete cycle of a solid body disk precession,  with the disk precession timescale rapidly running off to values exceeding the global evolutionary time of the disk. The corner of phase space which is best for observing precession is the limit $\alpha \to 0$ (which keeps the viscous time long despite the thick flow), $a\to 1$ (which makes the precession time as rapid as possible), and large $\beta$ and $M_\bullet$.  }
    \label{fig:chirp}
\end{figure*}

This is completely generic behavior for these solutions, as we show with a general chirp library in Figure \ref{fig:chirp}, which allows the five parameters ($M_\bullet, a, M_\star, \alpha$ and $\beta$) to vary (with fixed parameters of $M_\bullet = 10^6, M_\star = M_\odot, \alpha =0.03, a = 0.7, \beta = 1$ when not varied). We couple stellar radius and mass through $r_\star = m_\star^{4/5}$, and take a stripped mass of $\beta^3 m_\star/2$ when $\beta\leq 1$ (the partial TDE regime). The factor of $\beta^3$ for partial disruptions is informed by the simulations of \cite{Ryu+20a}. We see that for the vast majority of phase space one is unlikely to even see one complete cycle of a solid body disk precession (note the logarithmic time axis here),  with the disk precession timescale rapidly running off to values exceeding the global evolutionary time of the disk. 

\begin{figure*}
    \centering
    \includegraphics[width=0.49\linewidth]{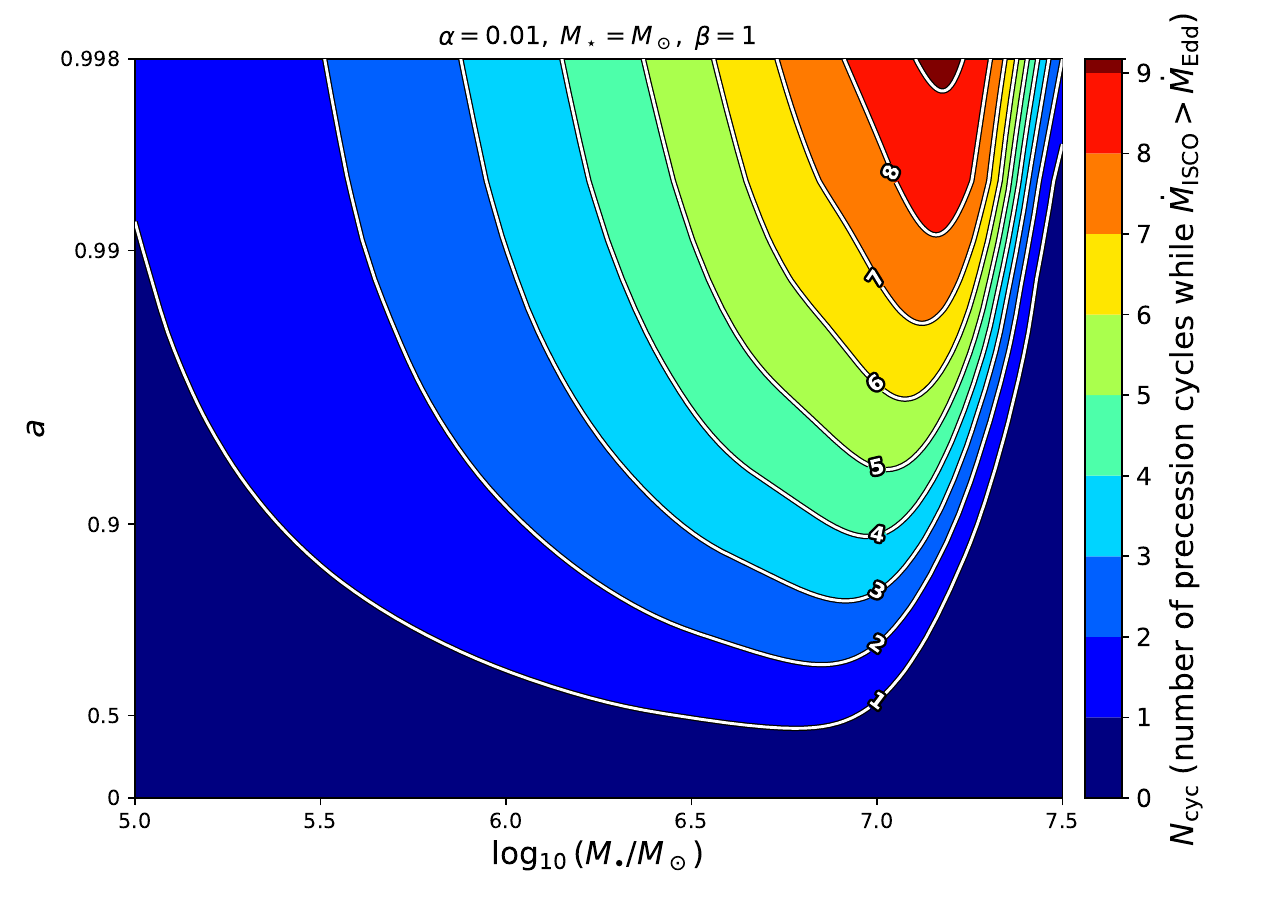}
    \includegraphics[width=0.49\linewidth]{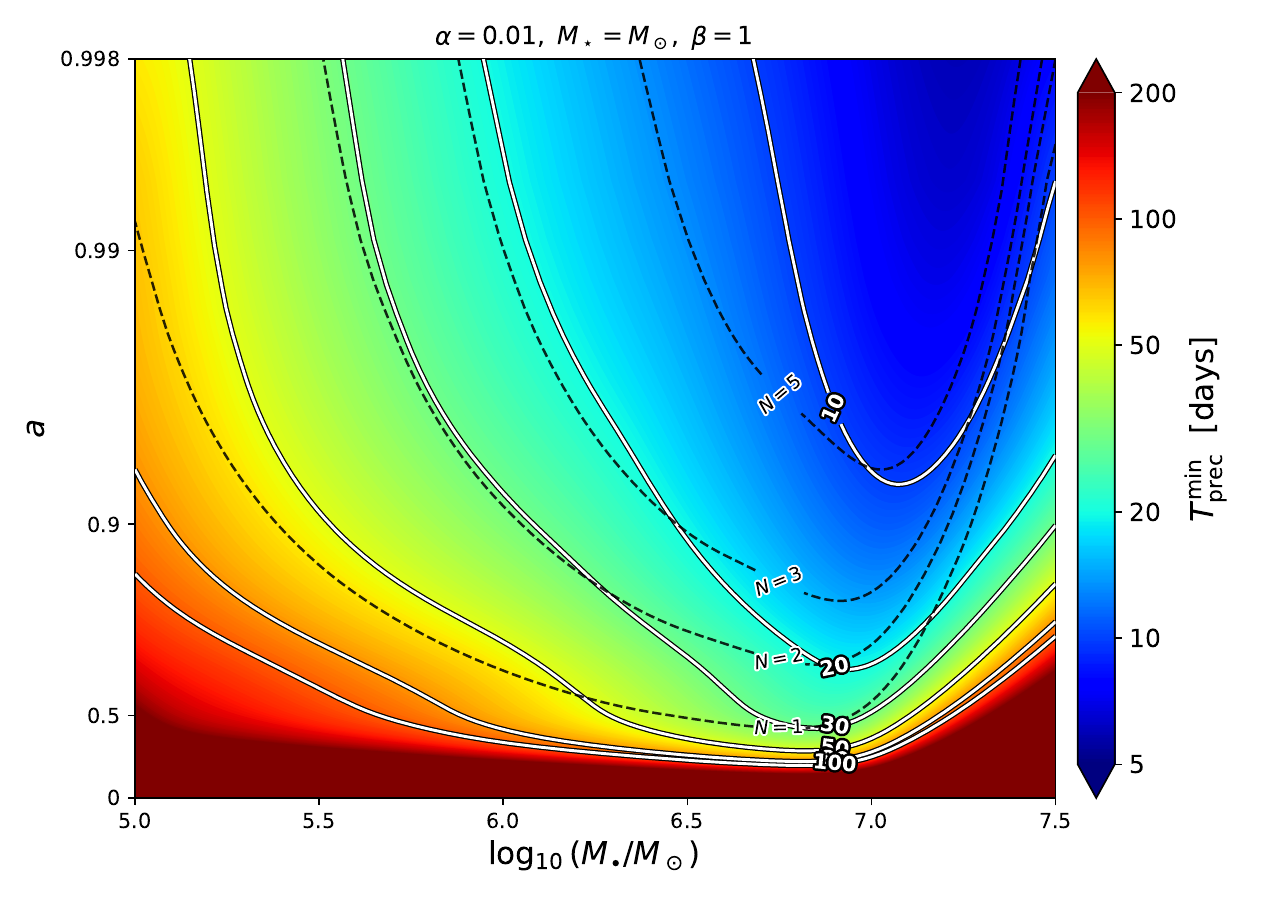}
    \includegraphics[width=0.49\linewidth]{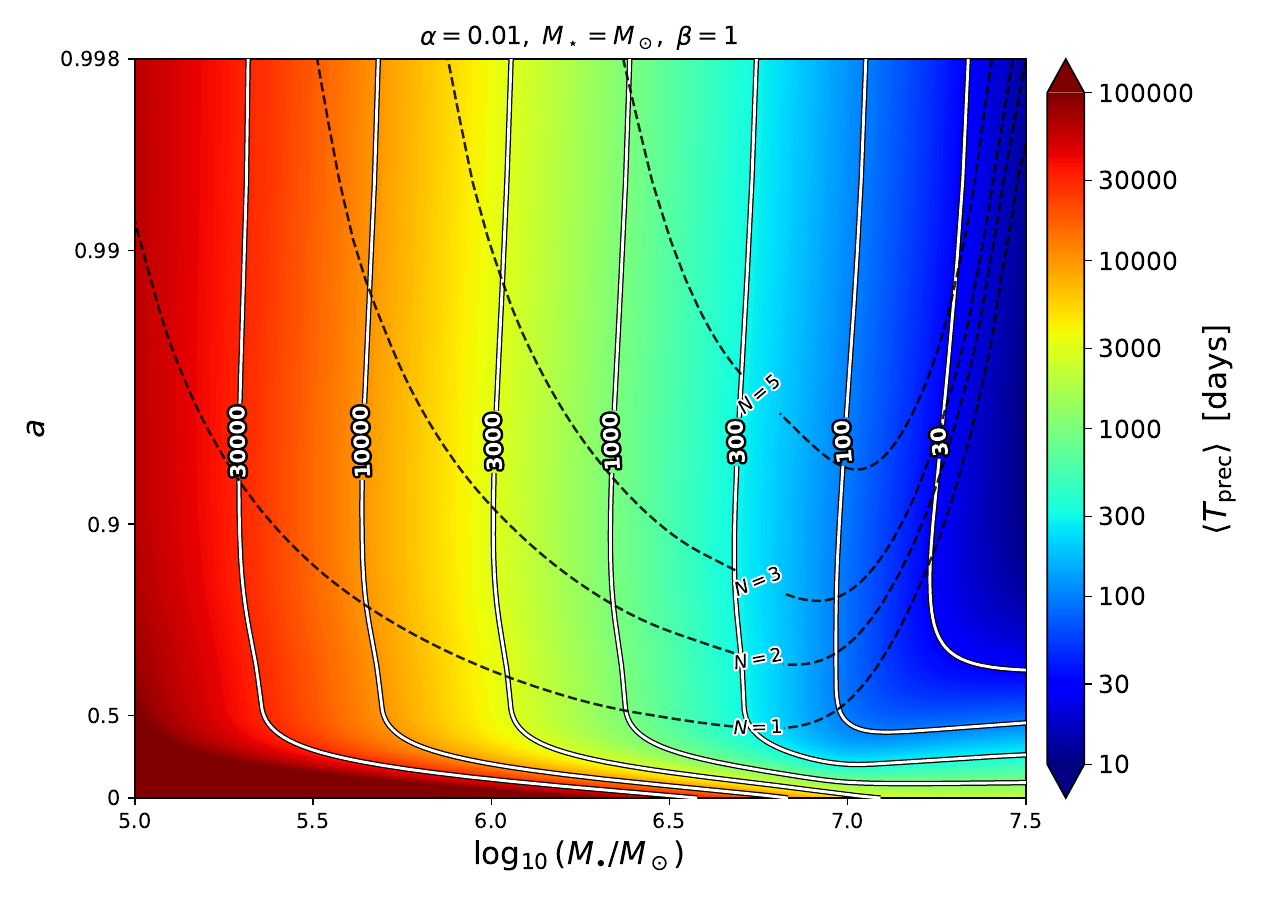}
    \includegraphics[width=0.49\linewidth]{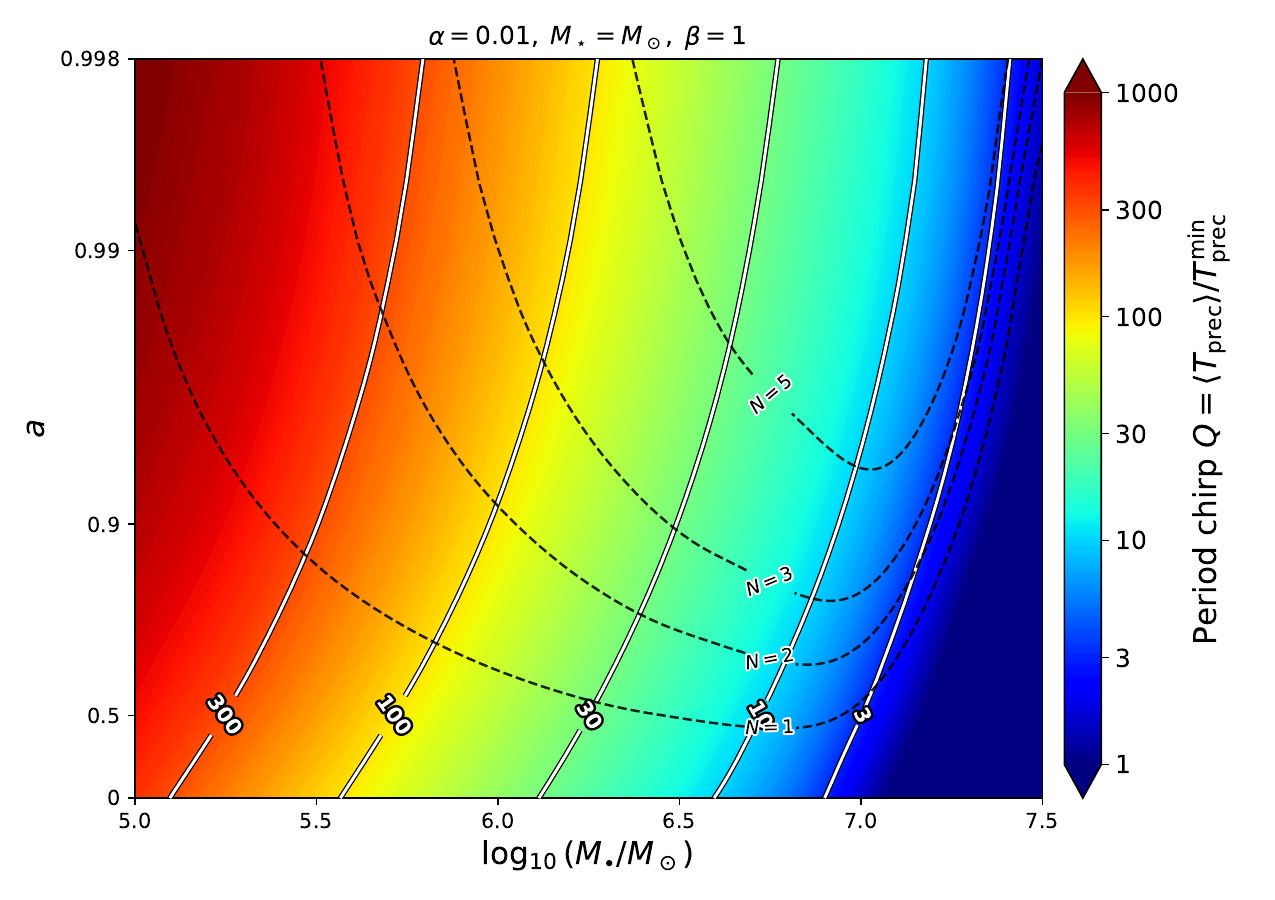}
    \caption{The region of black hole phase space $(M_\bullet, a)$ where one may see signs of precession in TDE data, for otherwise canonical TDE parameters $M_\star = M_\odot, \beta = 1, \alpha=0.01$. The top left panel shows the number of complete cycles (i.e., the number $N_{\rm cyc}$ of times $\phi$ passes through $2\pi$). We see that one only has an island of possible observability at $M_\bullet \sim 10^7 M_\odot$ for near maximal spin $a \sim 1$. This is because this region of phase space results in a compact initial disk which initially precesses rapidly. The other three panels show the initial precession period (upper right), the average precession period over the super-Eddington accretion phase (lower left) and the chirp (defined as the ratio of the average to initial precession periods). While much of TDE phase space results in initially observable precession periods (i.e., precession of order tens of days), the chirp is very strong at low black hole masses (lower panels), leading to no observable signal. The island of possible observability (high mass and spin) is also a region of low initial period and low chirp. The pole at $1/a$ for the minimum and average precession times can be seen by the colour bar saturation in the lower left and upper right panels.   }
    \label{fig:phase}
\end{figure*}

The corner of phase space which is best for observing precession is the limit $\alpha \to 0$ (which keeps the viscous time long despite the thick flow), $a\to 1$ (which makes the precession time as rapid as possible), $\beta \to \beta_{\rm max}$, $M_\bullet \to M_{\rm max}$ (both of which make the disk as compact initially as is possible, here $M_{\rm max}$ is the maximum black hole mass for a given star which produces super-Eddington fallback, and $\beta_{\rm max}$ is the maximum penetration parameter which keeps the initial orbit outside of the horizon). The stellar properties have minimal impact on the precession evolution. All of these results can be understood by inspection of the scaling analysis derived above
\begin{equation}
\frac{\Delta \Tprec}{\Tprec} \approx 10 \times   {\alpha r_I r_\star^{1/2} \over a M_6^{1/3} m_\star^{1/6} \beta^{1/2}}  . 
\end{equation}
Indeed, if there is to be any possibility of observing early-time precession in TDE X-ray emission, it will likely come from a small island of observability at high masses and (very) high spins. We show the black hole phase space of precession ($M_\bullet, a$) in Figure \ref{fig:phase}, where in the top left panel we show the number of completed cycles (i.e., how many times $\phi$ passes through $2\pi$) for otherwise canonical TDE parameters $M_\star = M_\odot, \beta = 1, \alpha=0.01$. We see that there is an island of possible observability at $M_\bullet \sim 10^7 M_\odot$ for near-maximal spin $a \sim 1$. This region of phase space results in a compact initial disk which initially precesses rapidly. The other three panels show the initial precession period (upper right), the average precession period over the super-Eddington accretion phase (lower left) and the chirp (defined as the ratio of the average to initial precession periods). While much of TDE phase space results in initially observable precession periods (i.e., precession of order tens of days), the chirp is very strong at low black hole masses (lower panels), leading to no observable signal. The island of possible observability (high mass and spin) is also a region of low initial period and low chirp, which aids in possible detectability. These disks are also those that only just pass through the Eddington mass accretion rate (the sharp cutoff in cycles in the uppermost right corner of the upper left panel is because the Eddington limit is never reached), and so are also the least likely to be perturbed by winds. 

\section{On the lack of global disk alignment}\label{sec:align}
The analysis in this paper thus far has been limited to the evolution of the global precession period of the flow. A second process competes with precession, namely disk alignment. Indeed, on sufficiently long timescales the disk inevitably aligns with the black hole spin (on the assumption that it continues to cool eternally and head towards $H/R\to 0$), but we shall show that this alignment cannot happen during the super-Eddington phase of the TDE disk except for highly fine tuned regions of parameter space. 

The physical origin of disk alignment is the small in-plane twist that the disk must develop in order to precess rigidly against a radius-dependent Lense-Thirring torque (each disk ring wants to precess at the local Lense-Thirring frequency, meaning that every ring must locally precess at the ``wrong'' frequency for there to be a single solid body precession period, this frequency mismatch induces a twist). Adjacent rings, twisted relative to each other, exert ``viscous'' stresses on one another, and the work done against these stresses dissipates the energy stored in the misalignment. We model this dissipation in the standard $\alpha$-prescription, with the understanding that it is ultimately  set instead by magnetohydrodynamic turbulence within the disk.

A first-order estimate of $t_{\rm align}$, which we will refine numerically below, follows from a simple energy budget argument. The disk twist stores an energy $E_{\rm twist}$ given by the work done by the residual Lense-Thirring torque against the angular twist of the disk. The disk's angular momentum $J_{\rm disk}$ sets the budget that this dissipation has to drain to bring the disk into alignment. The two combine into the alignment rate
\begin{equation}
    \frac{1}{t_{\rm align}} \sim\ \frac{E_{\rm twist}}{J_{\rm disk}} .
\end{equation}
The work done by the residual Lense-Thirring torque $\tau_{\rm LT}$ against the disk twist is, on dimensional grounds, $E_{\rm twist} \sim \tau_{\rm LT}\,\phi_{\rm max}$, where $\phi_{\rm max}$ is the characteristic twist angle across the entire disk. The total Lense-Thirring torque is, by definition of $\Omega_{\rm prec}$, $\tau_{\rm LT} \sim J_{\rm disk}\,\Omega_{\rm prec}$ (eq.~\ref{eq:Oprec_def}) and the residual torque has the same scale. Substituting into the heuristic alignment rate gives
\begin{equation}
    \frac{1}{t_{\rm align}} \sim \Omega_{\rm prec}\,\phi_{\rm max} \to  t_{\rm align} \sim \frac{T_{\rm prec}}{2\pi\,\phi_{\rm max}} .
\end{equation}
We see that $t_{\rm align}$ and $T_{\rm prec}$ chirp in direct proportion as the disk spreads, and so the same process that lengthens the precession period also lengthens the alignment time. 

The twist amplitude $\phi_{\rm max}$ is a quasi-steady state quantity which follows from steady-state viscous balance under the influence of a torque. The residual Lense-Thirring driving is resisted by a sound-wave-like restoring stress of order $\Sigma\,c_s^2 \sim \Sigma\,(H/R)^2\,R^2\,\Omega_K^2$, mediated by viscous coupling proportional to something like an $\alpha$ parameter. The result is heuristically 
\begin{equation}\label{eq:phimax_estimate}
    \phi_{\rm max} \sim \frac{\alpha}{(H/R)^2}\,\frac{\omega_{\rm LT}(R_{\rm in})}{\Omega(R_{\rm in})} \,,
\end{equation}
where each factor has the following physical origin: $\alpha$ because the twist exists only at first order in viscous coupling, $(H/R)^{-2}$ because a thick disk is a stiffer object which resists twisting, and $\omega_{\rm LT}/\Omega$ as the ratio of perturbing torque to unperturbed orbital frequency in the region where the disk torque is largest. Combining with the earlier results leads to 
\begin{equation}
    \frac{t_{\rm align}}{T_{\rm prec}} \;\sim\; \frac{(H/R)^2}{2\pi\,\alpha}\,\frac{\Omega(R_{\rm in})}{\omega_{\rm LT}(R_{\rm in})} \equiv N_{\rm crit} .
\end{equation}
This is a pure time-independent number (in the thick disk super-Eddington limit where $H/R$ is fixed) which has the following form  
\begin{equation}
    N_{\rm crit} \equiv {(H/R)^2 r_I^{3/2} \over 4\pi a \alpha},
\end{equation}
which is of order $\sim 10$ for high spins, and $\sim 100$'s for low spins. This result has a very important implication. Firstly, the disk alignment timescale is heavily chirped
\begin{equation}
    {\dot t_{\rm align} \over t_{\rm align}} =  {\dot T_{\rm prec} \over T_{\rm prec}} ,
\end{equation}
such that the disk does not rapidly align (as previous fixed-disk-size estimates have put forward). 

Indeed, generically, the disk never aligns during the super-Eddington phase. This is because the chirping of the alignment time means that global disk alignment is governed by a precession-cycle count rather than an explicit elapsed time, and the disk very rarely undergoes sufficient precession cycles. 

In the linear theory of warped disks in the wave like regime, the perpendicular component of the disks angular momentum (which is what causes the warp and precession in the first place)  is damped according to  ${\rm d}J_\perp/{\rm d}t = -J_\perp/t_{\rm align}(t)$ with $1/t_{\rm align}(t) = 1/(T_{\rm prec}(t) N_{\rm crit})$, which is trivially integrated to give
\begin{multline}
    J_\perp(t) = J_\perp(0)\,\exp\!\left[-\int{{\rm d}t' \over t_{\rm align}(t')}\right] \\ \approx J_\perp(0)\,\exp\!\left[-\frac{N_{\rm cycles}(t)}{N_{\rm crit}}\right] ,
\end{multline}
where 
\begin{equation}
N_{\rm cycles}(t) = \int_{t_{\rm fb}}^t {{\rm d}t' \over T_{\rm prec}(t')},
\end{equation} 
is the cumulative precession cycle count. The disk aligns by $1/e$ once it has completed $N_{\rm crit}$ cumulative precession cycles, regardless of how $T_{\rm prec}$ evolves between them. The disk therefore aligns within the super-Eddington phase only if $N_{\rm cycles}^{\rm SE} > N_{\rm crit}$, where $N_{\rm cycles}^{\rm SE}$ is the total number of cycles completed before the disk transitions out of the rigid-precession regime (see e.g., Figure \ref{fig:phase}).

\begin{figure*}
    \centering
    \includegraphics[width=0.95\linewidth]{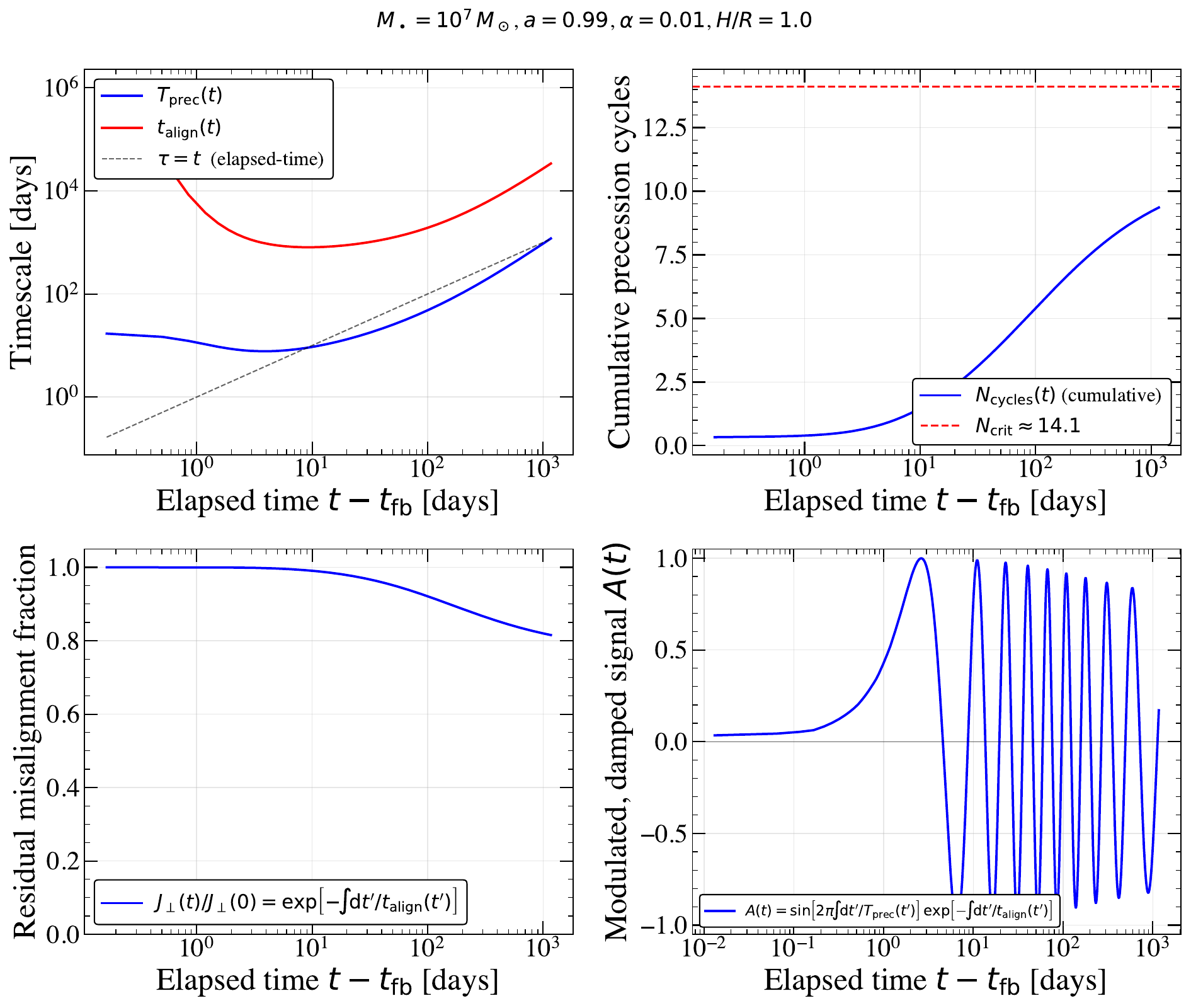}
    \caption{Upper left panel: the evolving precession (blue) and alignment (red) timescales of a flow with a high mass $M_\bullet = 10^7 M_\odot$, high spin $a=0.99$ and otherwise canonical TDE parameters $m_\star = 1$, $\alpha = 0.01$. We fix the viscous time by forcing $H/R = 1$. These parameters were chosen to be favorable for disk alignment. Even in this favorable limit the disk does not undergo enough precession cycles (upper right panel) to align, leaving a large residual misalignment even at $\sim 1000$ days post fallback (lower left panel). TDE disks will generically show global misalignment (and therefore warping) as they exit any super-Eddington phase (lower right panel).  }
    \label{fig:align}
\end{figure*}

\begin{figure*}
    \centering
    \includegraphics[width=0.48\linewidth]{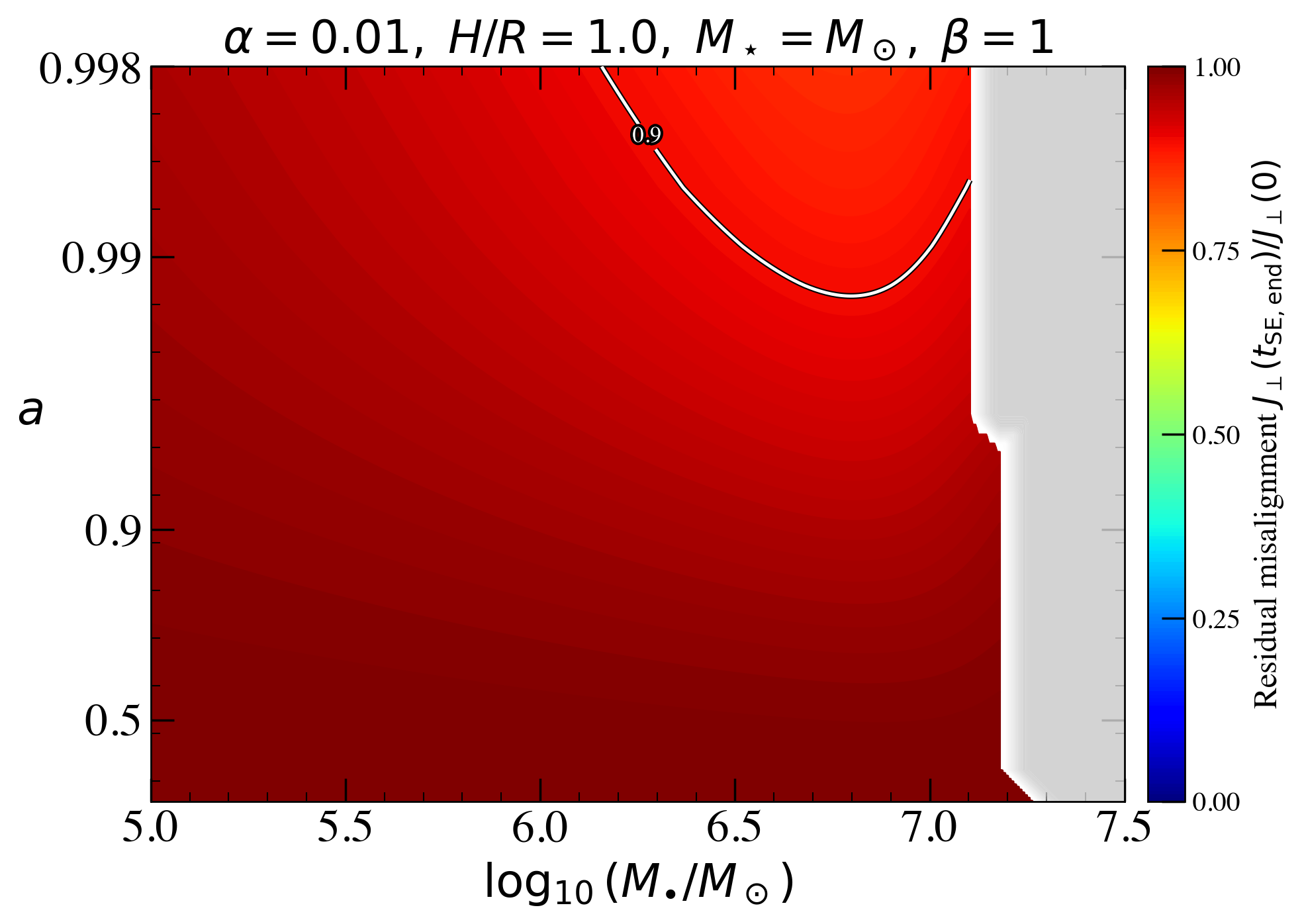}
    \includegraphics[width=0.48\linewidth]{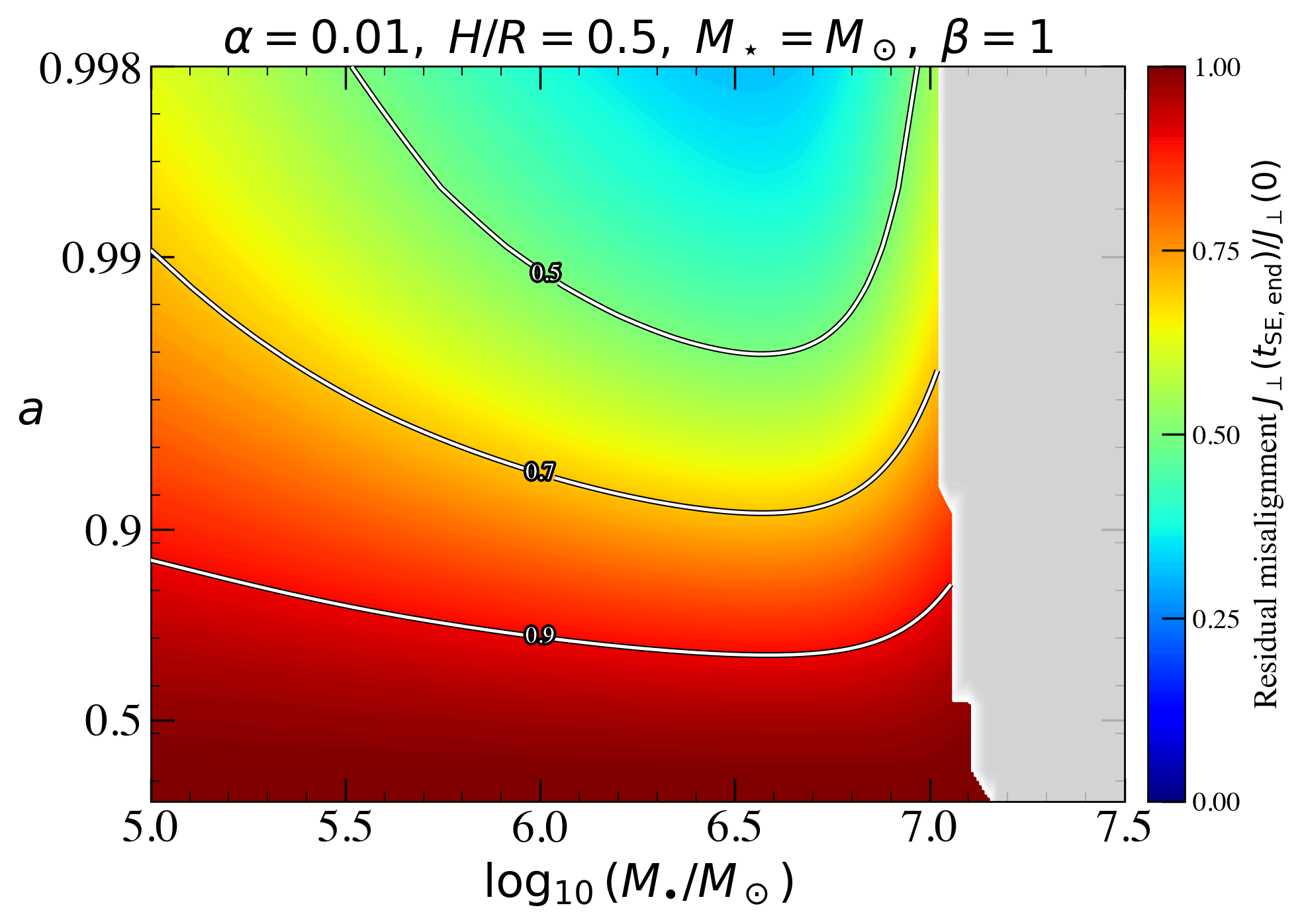}
    \includegraphics[width=0.48\linewidth]{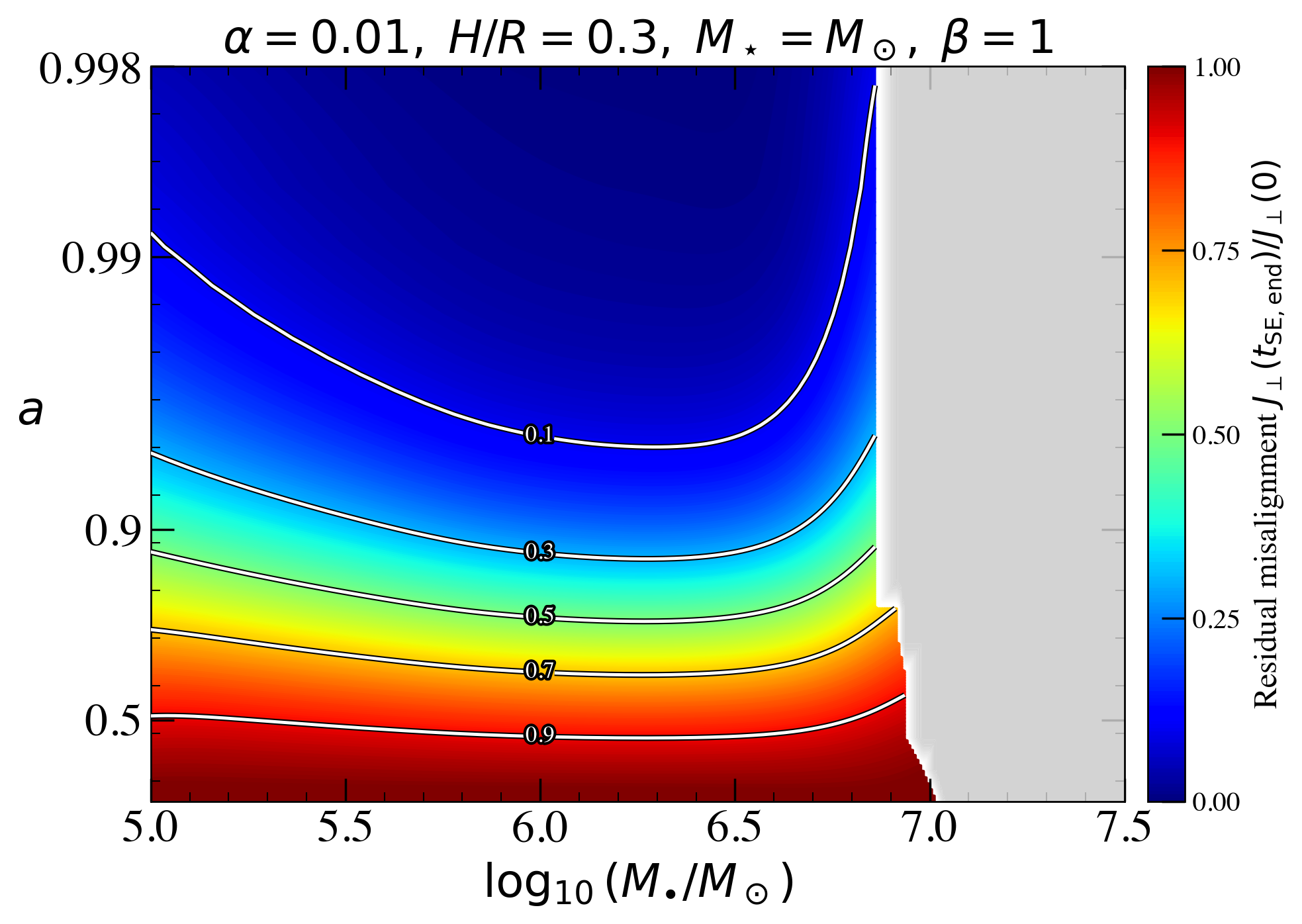}
    \includegraphics[width=0.48\linewidth]{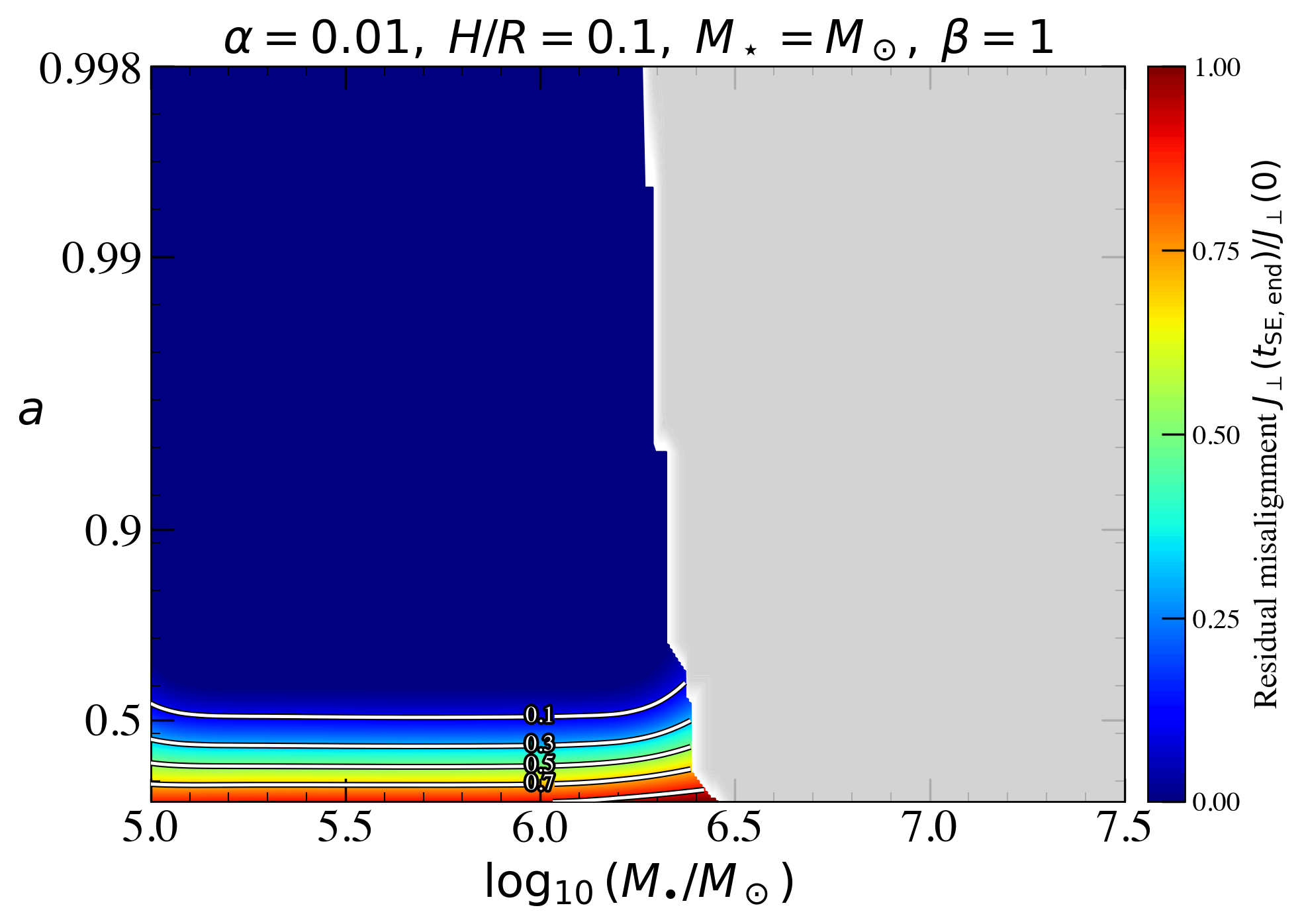}
    \caption{The residual angular momentum misalignment at the end of the TDE's super-Eddington phase, for different assumed disk aspect ratios at which a super-Eddington flow saturates. Grey regions of the phase space plot never reach a super-Eddington accretion rate. If super-Eddington accretion flows remain quite thin $H/R = 0.1$ (this would be surprising) then moderately rotating black holes $a\gtrsim 0.5$ would align (lower right panel), slightly thicker flows $H/R = 0.3$ barely align except above $a \gtrsim 0.9$, while thick $H/R\sim 0.5-1$ disks never align at any realistic spin.  }
    \label{fig:phase2}
\end{figure*}

For a disk experiencing the order-unity per-cycle period drift derived in Section~\ref{sec:DeltaT}, the cumulative cycle count grows only logarithmically with elapsed time. The same chirp that kills the solid body precession also limits $N_{\rm cycles}^{\rm SE}$ to a handful, generically less than $N_{\rm crit} \sim 10-100$. Indeed, an approximate condition for alignment is that the fractional precession period change per cycle is less than $1/N_{\rm crit}$, or 
\begin{equation}
    {\Delta T_{\rm prec} \over T_{\rm prec}} \lesssim {1\over N_{\rm crit}}, 
\end{equation}
which, as we had an analytical estimate of $\Delta T_{\rm prec}/T_{\rm prec}$ derived earlier can be reduced to a single inequality 
\begin{equation}
    {a^2 M_6^{2(1-p)/3} m_\star^{(1-p)/3} \beta^{1-p} \over r_I^{2+p} r_\star^{1-p}} > \left({H\over R}\right)^4 C(p),
\end{equation}
where $C(p) = (1+2p)94^{1-p}/8\pi$ is a constant. For our canonical $p=1/2$ model this reduces to $C(p) \approx 1$ and therefore 
\begin{equation}
    {a^2 M_6^{1/3} m_\star^{1/6} \beta^{1/2} \over r_I^{5/2} r_\star^{1/2}} > \left({H\over R}\right)^4. 
\end{equation}
Where we remind the reader that this alignment time is only valid in the thick-disk $H/R \sim 1 > \alpha$ wavelike regime (in other words, and importantly, it is not the correct condition when $H/R \ll 1$). Note that as well as being approximate, the above condition is an asymptotic condition (i.e., it assumes that the disk oscillates for infinite time), and a solution that satisfies the above may still not have enough time (during a super-Eddington phase) to align. 

To be more careful, we numerically compute the local disk precession timescale following the work of \cite{FoucartLai2014}. The alignment time is given by 
\begin{equation}
    \gamma_{\rm align} = {1\over t_{\rm align}} = {2\pi \int G_\phi \, {\rm d}\phi \over J_{\rm disk}} ,
\end{equation}
where $G_\phi$ is the residual Lense-Thirring torque, such that the work done (the energy budget to be dissipated) is the integral of this torque over the twist angle.

The residual torque in the model of \cite{FoucartLai2014} is given by the following integral
\begin{equation}
    G_\phi(R) \;=\; \int_{\rin}^{R}\Sigma\,R'^3\,\Omega\,\bigl[\Oprec - Z(R')\,\Omega(R')\bigr]\,{\rm d}R' ,
\end{equation}
with $Z(R) = (\Omega^2 - \Omega_z^2)/(2\Omega^2)$ a dimensionless function of the Kerr orbital and vertical epicyclic frequencies. In the limit $R \gg r_g$ this reduces to the familiar form $Z\,\Omega \to \wLT$. The precession frequency in this integral is computed as in the previous section. 
The in-plane twist gradient is \citep{FoucartLai2014}
\begin{equation}
{{\rm d}\phi\over {\rm d}R} = {4\alpha\,G_\phi\over\Sigma c_s^2 R^3}.
\end{equation}
We make the assumption that the equations of local hydrostatic balance can be expressed as $c_s^2 = (H/R)^2 v_K^2$, with constant $H/R$ to be specified. 
We evaluate the resulting $\gamma_{\rm align}(t)$ by computing $\Sigma(R,t)$ via the relativistic Greens function approach discussed earlier. As before the disk angular momentum is given by 
\begin{equation}
    J_{\rm disk} = 2\pi \int_{R_{\rm in}}^{R_{\rm out}} R' \Sigma U_\phi \, {\rm d}R'.
\end{equation}

We show an explicit numerical evolution of this integral in Figure \ref{fig:align}, which is for a region of parameter space most favorable for disk alignment (under a given assumption of $H/R=1$), namely high black hole mass $M_\bullet = 10^7 M_\odot$, and high spin $a=0.99$. We take $m_\star = 1$, and $\alpha = 0.01$ and fix the viscous time as before with $H/R = 1$. In the left hand panel we show the evolving time (dotted line), the evolving precession time (blue curve), and the evolving alignment time (red curve). One can see that initially the   disk alignment timescale looks like it  will drop to be short enough to naively align the disk during a super-Eddington phase (as has previously been argued to occur generically), but then strongly chirps to extremely large values. Even in this optimistic limit, the disk does not complete enough cycles to fully align (middle panel), meaning that there is still a large misalignment fraction $\sim \exp(-N_{\rm cycles}(t_{\rm SE})/N_{\rm crit})$ when the disk exits the super-Eddington phase (right panel). For lower spins, the disk barely aligns at even the $1\%$ level.  

The above result, and the degree of disk alignment generally, is sensitive to the value of disk thickness $H/R$ at which a highly super-Eddington accretion flows saturates (as can be seen in the above simple scaling analysis). This is in contrast with the precession result which is only sensitive to the product $\alpha (H/R)^2$. To examine this sensitivity we generate $(a, M_\bullet)$ phase space plots of the residual angular momentum misalignment fraction $J_\perp/J_{\perp, 0}$ at the time at which a super-Eddington phase of accretion ends for a given TDE system, under four different assumptions for the saturation scale height of a super-Eddington flow (flows which never reach a super-Eddington rate are shown in grey). The thinnest disk assumption ($H/R=0.1$, lower right panel) show that moderately rotating black holes $a\gtrsim 0.5$ would align, while slightly thicker flows $H/R = 0.3$ barely align except above $a \gtrsim 0.9$, while thick $H/R\sim 0.5-1$ disks never align at any realistic spin. While we do not claim to know precisely how the scale height of a disk will saturate in a highly super-Eddington flow, we think it unlikely to be as thin as $H/R \sim 0.1$, and therefore we anticipate large residual fractional misalignment across the entire TDE disk phase space. 

This result is relatively simple to understand, as a large fraction of the material which falls back is pushed out immediately, taking with it the angular momentum orientation of the incoming star (i.e., generically not aligned). The alignment time at large radii is very long, and the disk does not have time during the super-Eddington phase to catch up with the impacts of this expansion.

This means that -- at least at the start of -- observed thin-disk phases of TDE evolution one can expect the flow to exhibit a global warp structure. This is an interesting result which deserves further analysis, as it may well be the case that signatures of Lense-Thirring torques (and black hole spin) are imprinted in TDE observables, just not in the way which has been searched for (precession) thus far. 

\section{Implications, discussion and conclusions}\label{sec:con}
In this section we discuss some of the observational implications of our results for both TDEs and broader accretion timing phenomenology, and avenues for future theoretical investigation. 
\subsection{The observability of periodic modulation in X-rays from TDEs}
The most obvious implication of our work regards the general observability of signatures of Lense-Thirring precession in TDE disks, or more precisely, the general prediction of a lack of observability. This ``prediction'' is of course entirely consistent with the bulk of the TDE populations lack of any clear periodicity in the light curves \citep[e.g.,][]{Guolo24} or on short timescales \citep[e.g.,][]{Chakraborty26}. 

\begin{figure*}
    \centering
    \includegraphics[width=0.95\linewidth]{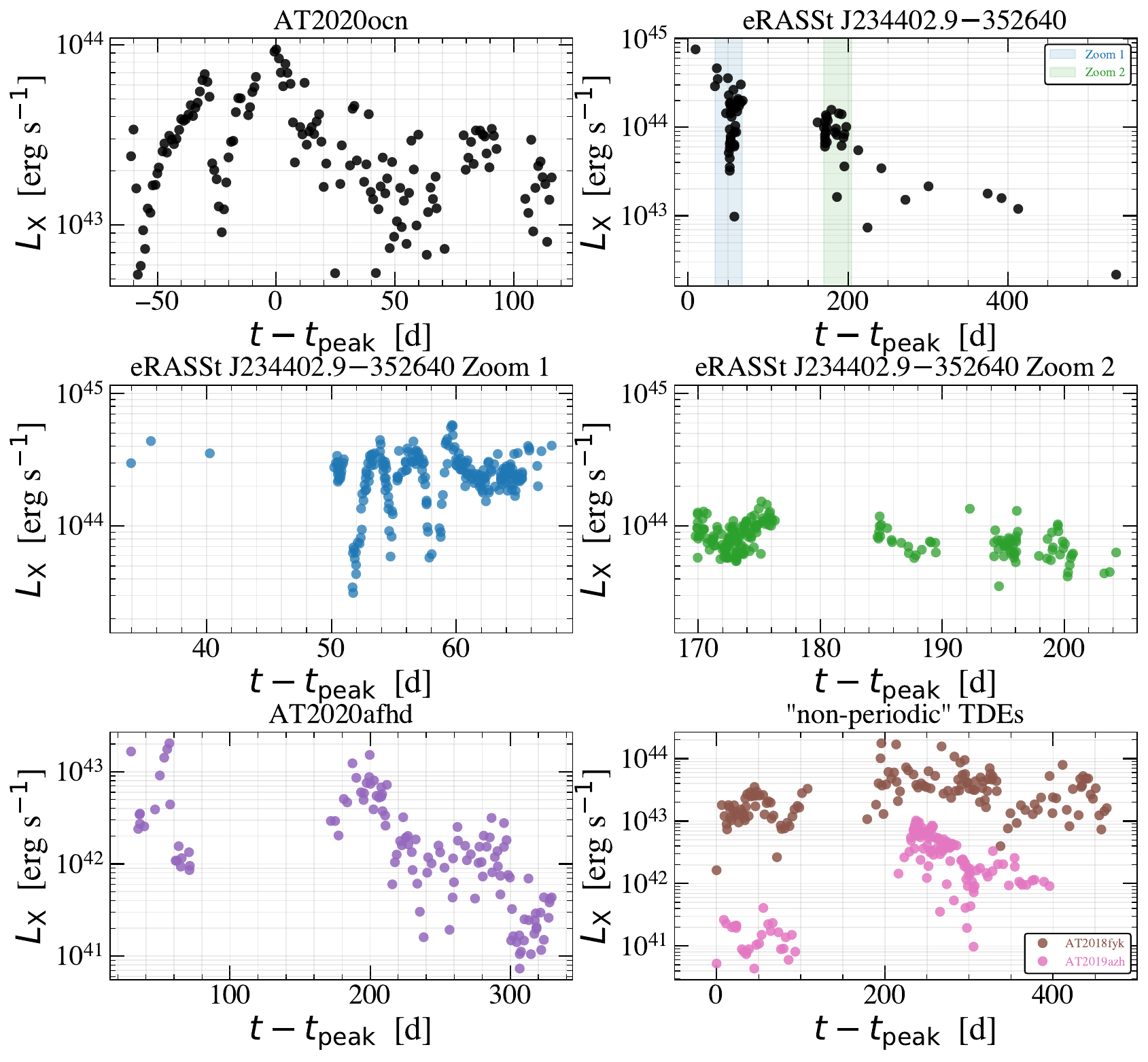}
    \caption{Claimed examples of Lense-Thirring precession seen in TDE X-ray light curves from the literature. The top left, top right, middle and lower left panels all show TDEs (and various zoom ins of stages in their evolution) which have claimed periodicity in their light curves, the bottom right highlights the sheer scale of the stochastic noise observed in ``non-periodic'' TDE X-ray  light curves.  The upper left example is AT2020ocn which shows $\sim 2$ dips during the rising phase of its light curve, followed by large amplitude noise on the decay, while eRASSt J2344 shows a brief period of $\sim 3$ dips during its rapid decline (lower left panel zooms in to the relevant section) and otherwise large amplitude short timescale noise. AT2020afhd looks, in the broader context of TDE X-ray variability (see AT2018fyk and AT2019azh in the lower right panel), like high amplitude noise, although was  claimed to also show periodic modulation in radio emission. The time offset $t_{\rm peak}$ here either refers to optical or X-ray peak, depending on how the source was discovered. }
    \label{fig:data}
\end{figure*}

There are three claims of periodic modulations in TDE X-ray light curves, and it is worth examining the data in detail. These are AT2020ocn \citep{Pasham24}, eRASSt J234402.9-352640 \citep{Malyali26} and AT2020afhd \citep{Wang25}. We show the raw X-ray light curves for all three in Fig \ref{fig:data} where we also include, for reference, the X-ray light curve of AT2019azh \citep{Hinkle21} and AT2018fyk \citep{Wevers19b} for which there is no claimed periodicity, but are included to highlight the sheer amplitude of multi-timescale noise which is present in TDE X-ray light curves. 

None of these events show what would be considered an unambiguous number of cycles ($\gg5$, say), with AT2020ocn showing $\sim 2$ dips during the rising phase of its light curve, followed by large amplitude noise on the decay, while eRASSt J2344 shows a brief period of $\sim 3$ dips during its rapid decline (lower left panel zooms in to the relevant section) and otherwise large amplitude short timescale noise. AT2020afhd looks, in context of TDE X-ray variability, like high amplitude noise, although was also claimed to show periodic modulation in radio emission (this radio variability is, however, also consistent with interstellar scintillation, A.J. Goodwin., priv. comm.). 

It is of course not possible to state with certainty whether either (i) the small number of apparent cycles seen truly reflect disk precession, rather than (for example) a stochastic disk turbulence \citep{MummeryTurner24, Mummery25Variability, Chakraborty26} process which briefly looks periodic, or (ii) whether the disappearance of these cycles reflects the realisation of the chirping mechanism presented here. It is probably worth noting that if Lense-Thirring precession did manifest in a TDEs X-ray light curve, it would be expected to look something like a small number of dips/flares ($\sim 1-3$) which rapidly lose coherence. 

This begs the question of whether it will ever be possible to robustly claim a detection of periodicity in a TDEs X-ray light curve. The false positive rate for a periodic signal as ``strong'' as that seen in AT2020ocn is quoted to be at the $\sim 10^{-4}$ level \citep[][]{Pasham24}. This may strike the reader as surprising when looking at the data itself, and indeed we find that the strength of this signal is strongly dependent on the null hypothesis made for the background noise. \cite{Pasham24} assume that the TDE light curve in the absence of any precession is described by a stationary stochastic process, which means that each null light curve is a damped random walk around a constant value (in other words the test for periodic signals is performed by comparing the light curve as seen to horizontal lines plus noise). This is an unusual choice for an astrophysical transient, and ultimately maximizes the size of the significance of the claimed periodicity (as secular evolution in the light curve itself gets absorbed into apparent periodicity signal as it cannot be successfully mimicked by a random walk). If one assumes there is a background evolution to the emission from the tidal disruption event itself (which seems like a reasonable model for a transient), and for example fits the light curve with a disk model \citep[e.g.][]{Mummery_et_al_2024}, then the significance of any periodicity drops significantly (and generally the most significant period shifts strongly as a function of ones choice for disk model parameters). Again, one is left with the possibility that the dips could be driven by Lense-Thirring precession, which is then chirped away, but there is no robust statistical model which can confirm or deny this, as $\sim 2-3$ cycles simply is not enough information given the underlying secular evolution of TDE accretion flows.

Finally, we note that it is not just thermal X-ray emission sourced from the inner disk which may plausibly be expected to show quasi-periodic modulation owing to solid-body precession. Indeed, jets/outflows which are launched along the disks rotation axis may also be anticipated to show some periodicity. Claimed $\sim$ periodicity in the literature includes the archetypal jetted TDE Swift J1644 \citep{Levan11}, which showed high amplitude X-ray (of non-thermal jet origin) variability at early times as well as the more recent noisy TDE AT2020afhd \citep{Wang25} which was argued to show $\sim$ periodicity in radio emission. 

Indeed, \cite{Teboul23} have argued that solid-body Lense-Thirring precession of TDE accretion flows may well explain the low incident rate ($\lesssim 1\%$) of jetted TDEs, as after a large number of precession periods the TDE environment will be encased in a dense ($\sim 4\pi$ solid angle) wind-like atmosphere through which the jet would struggle to escape. Our analysis differs from \cite{Teboul23} who pinned the disks (evolving) outer edge at a super-Eddington (trapping) radius, rather than the density edge we have used here. We believe it likely that either the entire disk will precess, or the flow will get stuck in a quasi-steady warp profile, and not that a sub-section of the disk can precess within a fixed outer disk. We acknowledge, however, that the evolution of warps in a disk where $H/R$ may vary significantly with radius (from the super-Eddington inner disk to the locally sub-Eddington outer disk) is a subtle problem that is not fully understood.   Our assumptions here lead to stronger precession period chirping compared to that of \citealt{Teboul23}, and are arguably more consistent with the lack of periodicity in TDE X-ray lightcurves.     It is not immediately obvious if the lower number of completed cycles in our framework impacts \citet{Teboul23}'s conclusions regarding jet formation.

\subsection{Precessing TDE disks as the origin of QPEs}
Another interesting phenomenology associated with TDE disks are quasi-periodic X-ray eruptions, or QPEs. These QPEs are soft X-ray flares observed from galactic nuclei evolving on the timescales of hours to days \citep[e.g.,][]{Miniutti2019,Giustini2020,Arcodia2021,Arcodia2024a,Chakraborty2021,Chakraborty25,Nicholl24,Hernandez25,Arcodia25ero5}. When in quiescence, the emission from an accretion disk is detected \citep{Nicholl24,Wevers25,Guolo25,Guolo25b} which, at least for some sources (and possibly all sources), is due to a previous tidal disruption event \citep[TDE; e.g.,][]{Quintin2023,Nicholl24,Chakraborty25,Guolo25b}. 

A recent model for the origin of quasi-periodic X-ray eruptions \citep{Middleton+2025:LT} postulated that QPEs are produced by a precessing accretion flow, and beaming of emission to the observer. Our analysis calls into question whether such a physical setup can be sustained for the length of time for which QPEs are observed in known systems. Indeed, eRO-QPE2 \citep{Arcodia2021} has shown a near constant $\sim 2$ hour period for $\gtrsim 5$ years \citep{Arcodia26} and RXJ1301.9+2747 \cite{Giustini2020} has shown a broadly constant period over $\sim 20$ years, both of which  would imply (within the model framework) a near constant precession period sustained by these disks over years-to-decades, something which is hard to reconcile with the strong chirp required by angular momentum conservation in the low mass black hole range of known QPE hosts. In effect one requires some mechanism to prevent the outer disk edge from expanding (to conserve angular momentum) while the flow is accreting at a sufficiently high rate to sustain a thick disk state (and therefore angular momentum is being lost inwards). 

One could imagine an inner region of the disk breaking off and precessing, with the much larger outer disk not undergoing any relativistic effects. This would keep the precession period short, but the same chirping should still apply to any isolated sub disk -- angular momentum conservation forces expansion at a rate coupled to accretion, and so any super-Eddington flow (with fixed mass budget not being externally fed at a constant rate) must expand rapidly and chirp away any periodic signal. A final problem for very long lived super-Eddington precession in a TDE disk is the finite mass budget $\dot M_{\rm acc} t_{\rm life}$ cannot exceed $f_\star M_\star$, where $f_\star$ encapsulates the messy problem of disk formation from the stellar debris. This bounds the lifetime of any super-Eddington phase to 
\begin{equation}
    t_{\rm life} < {f_\star M_\star \over f_{\rm Edd } \dot M_{\rm Edd}} \approx 50 \, {\rm years} \times \left({f_\star \over f_{\rm Edd} }\right) {m_\star \over M_6 } ,
\end{equation}
which for plausible values of disk formation efficiency $f_\star \sim 0.1$ and accretion rate  $f_{\rm Edd}\equiv \dot M/\dot M_{\rm Edd}\sim 10$ (required for a strong precession and beaming) are significantly shorter than known QPE lifetimes. The actual duration of super-Eddington fallback (plausibly bounding the period for which the accretion rate could be super-Eddington) 
\begin{equation}
    t_{\rm SE} \approx 1.5 \, {\rm years} \times  {m_\star^{1/5} r_\star^{3/5} \over M_6^{2/5} } ,
\end{equation}
is again uncomfortably short when compared to QPE lifetimes. 

\subsection{The impact of disk precession on QPE timing}
In an alternative model for the origin of QPEs, the flares themselves are produced by collisions between some orbiting body and the accretion flow \citep{Xian2021, Linial2023, Franchini2023, Yao24, Mummery25QPE}. The timing structure seen in QPE flares is then assumed to track the times at which the orbiter crosses the disk plane. The possibility that the TDE accretion disk might precess is left as a free parameter in many timing fitting codes \citep{Franchini2023,Chakraborty25b, Miniutti25}. This then opens up a broad timing phenomenology, as one can get very simply get beating between the orbiter's period and an invoked disk precession period.  While this naturally opens up interesting timing phenomenology, we do not believe it is justified physically on a number of grounds. Firstly, the natural precession period for the low mass black holes in QPE sources runs away to extreme values $\sim$ years, making it unlikely to impact any observable timing features on the short timescales probed by dedicated observing campaigns of order $\sim$ weeks-months (though could of course impact the evolution over year timescales). 

Secondly, and perhaps more pertinently, QPEs are found in the late time stage of TDEs, where the disk is well modeled by a thin, sub-Eddington flow, explicitly not the physical limit in which disk precession is expected. Instead, the disk is more likely to be in the diffusive ($H/R \lesssim \alpha$) and warped regime, the more natural route for searching for signs of relativistic (Lense-Thirring torques) is through the long-term settling of the disk state into a warped structure. Our results, particularly  the lack of disk alignment with the black hole spin axis mean, if orbiters are indeed the origin of (some) QPE flares, that this orbiter will be interacting with a disk structure which is fundamentally not two dimensional. While we do not believe disk precession will impact timing signatures of orbiter-driven QPEs on short timescales, this global warp may well (the nodal precession of the orbiter will interact with the $>2$ dimensional structure of the disk to induce non-trivial timing features, for example). This possibility deserves future scrutiny and will be the focus of follow up work. 

\subsection{A better model moving forward}
Our analysis in this paper has been limited to a simplified model for the Lense-Thirring driven precession and alignment of the disk, namely we have made the  assumption of pure solid body precession while forcing (by hand) a thick disk aspect ratio. The purpose of this was to highlight and examine a key physical point missing in earlier analyses of the physics of this regime, namely that the angular-momentum-conservation enforced rapid expansion of the flow leads to strong chirping of the disk precession and alignment timescales. There are however other effects one would wish to include in a more detailed model of this early evolution. 

A better model would solve the full set of linear warp equations in the wavelike regime \citep[a subset of these linear warp equations was solved by][]{Franchini2016}, with these linear warp equations coupled to the disk surface density evolution equation (so that radial expansion can be coupled to the warp/precession evolution). Further coupled to this should be a dynamical equation tracking the local disk scale height $H/R$, whose value determines whether or not the disk responds diffusively or in a wave-like manner to the Lense-Thirring torque. This would then more carefully track both the chirping precession period, the expansion of the disk, and impacts of disk alignment (as the flow thins) and the possible onset of quasi-steady-state disk warping. Such work is of great interest, both in its more complete description of the evolution of Lense-Thirring precession, but also for its predictions regarding the final state of the flow and any large scale warp dynamics. 

\subsection{Signatures of Lense-Thirring torques -- warped disks}
An important result of our analysis is the prediction that, rather than generically showing signatures of Lense-Thirring precession (as was previously assumed), TDE accretion flows should instead generically show signs of quasi-steady-state Lense-Thirring warps, at least when they first drop out of the super-Eddington accretion phase. 

This simply results from the fact that the material which first falls back inherits the specific angular momentum of the stellar orbit, and is almost immediately pushed out to large radii (this is a consequence of rapid early accretion).  This material, as we showed in section \ref{sec:align}, never has the opportunity to align with the black holes spin axis during any possible super-Eddington phase. This means that as the disk enters the thin disk phase and has warp evolution which is more likely governed by a diffusive response rather than a wavelike response, it will have material at large scales which is highly misaligned. In the diffusive regime the inner disk will presumably relatively quickly align with the spin axis, but this means that the disk will exhibit a global warp, at least initially. Possible observational signatures of such a global warp are of interest as a probe of relativistic physics, and will be the focus of a follow-up paper.

\subsection{Conclusions}
In this paper we have examined the impacts of chirping on the evolution of Lense-Thirring precession and global disk alignment of purported thick TDE accretion disks. The impacts of angular momentum conservation and disk spreading (required for accretion) mean that the vast majority of TDE phase space will rapidly exit a region of fast coherent precession, making it rather unlikely to ever be detected for more than $\sim 3$ cycles at best.  We argue that this is consistent with existing X-ray data on TDEs. Global disk alignment is equally strongly impacted, and is unlikely to occur for canonical regions of TDE parameter space.  Models of QPE timing which invoke the precession of TDE disks, either as the origin of QPEs, or to explain anomalous timing features of orbiting models, will likely have to be revisited in light of the results found here.

The lack of disk alignment during the wavelike regime means, generically, that TDE disks will enter their thin disk stage of evolution with a global warp. It is likely therefore that signatures of relativistic Lense-Thirring physics are to be found in this later, more well behaved, stage of evolution. Late stage TDE disks therefore represent exciting probes of strong gravity physics, if observational diagnostics of these warps can be determined. 

{Throughout this paper we have intentionally focused on the initial super-Eddington phase of TDE disk accretion.    As the disk spreads and the accretion rate decreases the disk will transition to a thin radiation pressure supported disk, for which the structure and magnetic stress prescription is less well understood (e.g., \citealt{huang_global_2023,Zhang2026}).   Constraints on warps in late stage TDE disks could thus also provide a potentially powerful  probe of the geometric thickness and magnetic stress in thin radiation pressure dominated disks.}

\section*{Acknowledgments}
A.M. is grateful to Callum Fairbairn for various stimulating conversations throughout the completion of this work. A.M. acknowledges support from the Ambrose Monell Foundation, the W.M. Keck Foundation and the John N. Bahcall Fellowship Fund at the Institute for Advanced Study. 

\label{lastpage}

\bibliography{andy}
\bibliographystyle{aasjournal}

\end{document}